\documentclass[floats,floatfix, amssymb, prd,nofootinbib,superscriptaddress]{revtex4-2}

\usepackage{amssymb,amsmath,verbatim,mathtools,needspace,enumitem,etoolbox,graphicx,physics,microtype,afterpage,xspace,tabularx,lmodern,multirow}
\usepackage{subcaption} 
\usepackage{gensymb}
\usepackage[dvipsnames, usenames]{xcolor}
\usepackage[unicode, colorlinks=true, linkcolor=linkcolor, citecolor=linkcolor, filecolor=linkcolor, urlcolor=linkcolor, linktocpage, breaklinks]{hyperref}
\usepackage[nolist,nohyperlinks]{acronym}
\usepackage{tikz}

\newcommand{\orcid}[1]{\href{https://orcid.org/#1}{\includegraphics[width=10pt]{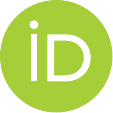}}}

\definecolor{linkcolor}{rgb}{0.6, 0, 0.5}

\usetikzlibrary{decorations.markings}
\usetikzlibrary{decorations.pathmorphing}

\allowdisplaybreaks

\makeatletter 
\renewcommand\onecolumngrid{
\do@columngrid{one}{\@ne}
\def\set@footnotewidth{\onecolumngrid}
\def\footnoterule{\kern-6pt\hrule width 1.5in\kern6pt}
}
\makeatother

\begin{document}

\pagenumbering{arabic}

\title{Post-Minkowskian expansion of the Prompt Response in a Schwarzschild background}

\author{Marina De Amicis \orcid{0000-0003-0808-3026}}
\email{mdeamicis@perimeterinstitute.ca}
\affiliation{Perimeter Institute for Theoretical Physics, 31 Caroline St N, Waterloo, ON N2L 2Y5, Canada}
%
\author{Enrico Cannizzaro \orcid{0000-0002-9109-0675}}
\email{enrico.cannizzaro@tecnico.ulisboa.pt}
\affiliation{CENTRA, Departamento de Fisica, Instituto Superior T\'ecnico – IST,
Universidade de Lisboa – UL, Avenida Rovisco Pais 1, 1049 Lisboa, Portugal}

\begin{abstract}

We study the early-time component of the Green's function of a Schwarzschild black hole, traveling on the curved light cone and usually denoted as the prompt response.
Working in a post-Minkowskian approximation, we show for the first time that the prompt response is given by the residue of poles at $\omega=0$ present in the complex Fourier domain.
The contribution of the high-frequency arcs, previously assumed to generate the prompt response, vanishes.
The analytical expression of the prompt response in this scheme is a polynomial of order $\ell$ in the observer's retarded time, with $\ell$ the multipole number.
We validate the model against numerical predictions, obtaining good agreement for a compact source far from the black hole. We provide a phenomenologically-corrected expression to improve the match as the source is moved closer.
We investigate the polynomial structure of the prompt response for sources close to the black hole through a series of numerical fits.
Our work is a fundamental step in the broader effort to develop first-principles, analytical models for binary black hole coalescence signals, valid close to the merger and during the early ringdown stage.

\end{abstract}

\maketitle


\section{Introduction} %
%

%
Gravitational waves (GWs) from coalescing compact objects are now routinely observed by the LIGO–Virgo–KAGRA network~\cite{LIGOScientific:2014pky,VIRGO:2014yos,KAGRA:2020tym}, with steadily improving detector sensitivity~\cite{LIGOScientific:2025hdt}, and new detectors planned or in construction~\cite{Saleem:2021iwi,Pandey:2024mlo, Colpi:2024xhw,Abac:2025saz}. These observations provide access to the dynamical, strong field regime of gravity and enable precision measurements of the properties of compact binaries, their formation channels, and the nature of gravity itself. In particular, binary black hole (BH) mergers offer a unique laboratory for testing general relativity predictions (GR) and to probe the existence of new fundamental fields~\cite{Berti:2015itd,LIGOScientific:2021sio,LIGOScientific:2025rsn,LIGOScientific:2025rid,LIGOScientific:2025brd,LIGOScientific:2025csr}, or to constrain exotic compact objects~\cite{Cardoso:2019rvt}.
In order to extract information from GWs observations, signal modeling is required. Typically, black-hole binary mergers are divided into three main stages: an inspiral, during which the two BHs are orbiting each other on a (slowly) decaying orbit; a plunge-merger phase, which includes the last stage of the two-body problem process (plunge) right before a common horizon has formed (merger); a stationary ringdown regime, which can be treated perturbatively (mainly) as a superposition of exponentially-damped oscillations, denoted as quasinormal modes, with constant amplitudes.
Numerical relativity (NR)~\cite{10.1093/acprof:oso/9780199205677.001.0001} provides accurate waveforms to describe the whole black-hole binary evolution up to late times. However, NR simulations have some limitations, e.g., being computationally expensive and not suited for agnostic tests of GR.
Hence, it is useful to complement analyses carried through NR models with others performed through analytical, ``closed-form'' ones, which can either be purely phenomenological or derived by first principles (i.e. solving Einstein's equations in certain approximation schemes). Phenomenological models are more limited than first-principles ones since they are not written in terms of observables with immediate physical interpretation.
Analytical first-principles models are available during the inspiral (through the self-force approach~\cite{Pound:2021qin} and the effective one body formalism~\cite{Buonanno:1998gg,Damour:2009zoi}) and for the stationary ringdown stage, which begins $\sim 10-20M$~\cite{London:2014cma,London:2018gaq} after the quadrupole peak in luminosity, generally considered close to the merger time.
The ``transient'' signal emitted in the plunge-merger stage is described, instead, through phenomenological models~\cite{Baker:2008mj,Damour:2014yha}. 
Curiously, such models have the same functional form for both the extreme mass-ratio limit and for comparable mass mergers.

The lack of a plunge-merger analytical model, even in the linear, test-particle, non-spinning limit, is due to a limited knowledge of a close-form expression for the Schwarzschild Green's function (GF).
Leaver pioneered the investigation of this GF, and in Ref.~\cite{Leaver:1986gd} he identified its three different components: an initial signal traveling on the curved light-cone and denoted as prompt response; a superposition of exponentially damped vibrational modes, called quasinormal modes (QNMs) (see also Ref.~\cite{Leaver:1985ax}) originated from the back-scattering off the peak of the background potential barrier; and an inverse power-law decay denoted tail (first found in Ref.~\cite{Price:1971fb}), due to the interaction with the long-range background curvature.
Recently, Ref.~\cite{DeAmicis:2025xuh} extended part of this picture to the test-particle limit, introducing a notion of causality for the QNMs propagation, allowing to track the QNMs excitation during the plunge-merger stage. 
Ref.~\cite{DeAmicis:2025xuh} shows that other components of the Green's function are needed to model the transient leading to the stationary ringdown (see Fig.~8 of said work).
The tail is mainly excited at large distances and appears to be sub-leading with respect of both the prompt response and the QNMs, which is the reason why it only dominates the signal at late times. 
Hence, we speculate that the missing piece in the analytical reconstruction of the plunge-merger signal in Ref.~\cite{DeAmicis:2025xuh}, is due to neglecting the prompt response.
Past work on modeling the prompt response signal in Schwarzschild is limited to the heuristical intuition of Leaver~\cite{Leaver:1986gd} and a computation of Andersson~\cite{Andersson:1996cm} valid for a scalar field. The latter starts from a high-frequency approximation, to later show that the prompt response is, in the frequency domain, the Laplace transform of an Heaviside function (i.e. a small frequency effect). 
Recent literature~\cite{Arnaudo:2025uos,Kuntz:2025gdq} discussed the early-time response for other geometries, namely Schwarzschild-de Sitter and P$\mathrm{\ddot{o}}$schl-Teller, showing that it comes from poles of the Laplace-transformed GF on the imaginary-frequency axis.
However, an analytical expression for the prompt response in Schwarzschild is still missing.
Another attempt to analyze the transient near the peak which is worth discussing was performed in Ref.~\cite{Oshita:2025qmn}. In this work, the so-called ``direct wave'' signal is studied, arguing that it is the component of the signal which directly travels from the source to the observer, without back-scattering from the potential barrier peak. We stress that this signal is obtained from features of the GF in the complex-frequency plane which are different from the ones originating the time-domain prompt response, as studied in the present work.
Moreover, the contour integral used in Ref.~\cite{Oshita:2025qmn} to anti-transform back to the time domain is not the one originally proposed in Ref.~\cite{Leaver:1986gd}, so that it is not clear if there is an identity transformation between ours or Leaver's GF decompositions, and that of Ref.~\cite{Oshita:2025qmn}, hence between the prompt response and the direct wave.

In the present work, we provide an analytical expression of the prompt response in a post-Minkowskian approximation scheme, assuming compact sources localized at large distances, which excite the small frequency response of the background. 
We find that, in this limit, the prompt response of a Schwarzschild background comes from poles on the imaginary axis at $\omega=0$, similar to what is observed in Ref.~\cite{Arnaudo:2025uos,Kuntz:2025gdq} for the Schwarzschild-de Sitter and P$\mathrm{\ddot{o}}$schl-Teller geometries.
The residues of the poles give rise to a polynomial behavior of order $\ell$, with $\ell$ waveform multipole, in the retarded time of an observer at null infinity $(\mathcal{I}^+)$.
We validate this prediction against numerical simulations, showing good agreement for sources at large distances. 
For sources closer to the BH, the predictions show a progressing dephasing with respect to the numerical prompt response, that is solved introducing an heuristically-corrected model. 
While at intermediate and small distances our model fails (for sources $\lesssim100 M$), we show through a series of fits on the numerical waveform that the prompt response is still partially described by a polynomial of order $\ell$ in the observed retarded time.
The paper is structured as follows. 
In Sec.~\ref{sec:background} we introduce the notation and framework used in this work, along with some GF generalities, and information on the private code \textsc{RWZHyp} through which the numerical evolutions have been computed.
In Sec.~\ref{sec:flat_spacetime} we compute the time-domain prompt response in the asymptotic flat spacetime limit, and show it originates as residue of poles at $\omega=0$ in the complex-frequency domain.
In Sec.~\ref{sec:Schwarzschild}, we extend this computation to a Schwarzschild background leveraging a post-Minkowskian approximation scheme, with corrections up to $\mathcal{O}[(M/r')^3]$, with $r'$ location of the impulsive source.
In Sec.~\ref{sec:results}, we compare the predictions computed in Sec.~\ref{sec:Schwarzschild} against numerical experiments, we provide a semi-phenomenological corrected expression of the prompt response valid as the source is progressively closer to the black hole, and we investigate the nature of this signal when excited by a source at small distances through a series of numerical fits.
We summarize and provide insights for future directions in Sec.~\ref{sec:conclusions}. 
%

\section{Framework}   %
\label{sec:background} %
%

We consider small perturbations on top of a fixed Schwarzschild geometry, whose line element is
\begin{equation}
ds^2=-A(r)dt^2+\frac{dr^2}{A(r)}+r^2d\Omega^2 \ , \  A(r)=1-\frac{2M}{r} \, ,
\label{eq:Schwarzschild_metric}
\end{equation}
Given the background symmetry, the perturbations can be decomposed into spin-2 weight spherical harmonics ${}_{-2}Y_{\ell m}$
\begin{equation}
    h(t,r,\Theta,\Phi) =  \sum_{\ell \geq 2,|m|\leq \ell} h_{\ell m}(t,r) _{-2}Y_{\ell m}(\Theta,\Phi)
\end{equation}
We can identify two sectors, according to how the perturbations transforms under parity: the even sector and the odd one. Even (odd) perturbations are eigenstates of the parity operator with eigenvalue given by $(-1)^{\ell}$ ($(-1)^{\ell+1}$).
For each sector, it is possible to build a gauge invariant master function: the Regge-Wheeler one for odd perturbations, $\Psi_{\ell m}^{o}$, and the Zerilli one for even ones, $\Psi_{\ell m}^{e}$.
These master functions satisfy Schroedinger-like equations of the form
\begin{equation}
\left[\partial_t^2 -\partial_{r_*}^2+V^{e/o}_{\ell m}(r_*)\right]\Psi_{\ell m}^{e/o}(t,r_*) = 0 \, ,
\label{eq:RWZ_equation}
\end{equation}
where we have introduced the tortoise coordinate with the convention $r_*=r+2M \, \log\left(r/(2M)-1\right)$.
At large distances, the master functions are related to the strain multipoles through the following expression~\cite{Nagar:2005ea}
\begin{equation}
h_{\ell m}=\dfrac{M}{r}\sqrt{\dfrac{\left(\ell+2\right)!}{\left(\ell-2\right)!}}\left(\Psi^{e}_{\ell m}+i \Psi^{o}_{\ell m}\right)+\mathcal{O}\left(\dfrac{M^2}{r^2}\right) \,,
\end{equation}
where, depending on the parity, only one of the two terms on the right-hand side vanishes.
We focus on initial-data perturbations, in particular, setting 
\begin{equation}
\Psi_{\ell m}(t=t',r_*)=0 \ \ , \ \ \ \ \, \partial_t \Psi_{\ell m}(t=t',r_*)= \frac{1}{\sqrt{2\pi}a}\exp{\left[-\frac{(r_*-r_{*}')^2}{2a^2}\right]}\, .
\label{eq:initial_data_gauss}
\end{equation}
The initial-value problem specified by  Eq.~\eqref{eq:RWZ_equation} and Eq.~\eqref{eq:initial_data_gauss} is equivalent to solving the inhomogeneous equation
\begin{equation}
\left[\partial_t^2 -\partial_{r_*}^2+V^{e/o}_{\ell m}(r_*)\right]\Psi_{\ell m}(t,r_*) = \mathcal{S}_{\ell m}(t,r_*) \, ,
\label{eq:RWZ_equation_sourced}
\end{equation}
with a particular time dependence of the source, given by
\begin{equation}
\begin{split}
S^{\rm ID}_{\ell m}(t,r_*)=\Psi_{\ell m}(t',r_*)\partial_t\delta(t-t')+
\partial_t\Psi_{\ell m}(t',r_*)\delta(t-t') \, .
\end{split}
\label{eq:ID_source}
\end{equation}
For an infinitely sharp gaussian, in the limit $a\rightarrow 0$, the above problem reduces to the definition of the  Green's function
\begin{equation}
\left[\partial_t^2 -\partial_{r_*}^2+V^{e/o}_{\ell m}(r_*)\right]G_{\ell m}(t-t';r_*,r_{*}') = \delta(t-t')\delta(r_*-r_{*}') \, .
\label{eq:GF_definition}
\end{equation}
We will present numerical investigations of the Green's function in the time domain as observed at $\mathcal{I}^+$, obtained with the \textsc{RWZHyp} code~\cite{Bernuzzi:2011aj, Bernuzzi:2010ty}.
The software \textsc{RWZHyp} features a double-precision implementation and uses a homogeneous grid in $r_*$ that for negative radii ends at a finite value, in order to keep the horizon outside of the computational domain.
At large distances, a hyperboloidal layer is attached to the computational domain, allowing to solve for the gravitational perturbation $\Psi_{\ell m}$ at future null infinity $\mathcal{I}^+$.
The coordinates of the layer are the retarded time $\tau$ and the compactified spatial coordinate $\rho$. We refer to~\cite{Bernuzzi:2011aj} for an explicit expression of the coordinate $\rho$ in terms of $(t,r_*)$, while for $\tau$ it holds
\begin{equation}
    \tau-\rho=t-r_* \, .
    \label{eq:RWZcoordinates}
\end{equation}
We use the same resolution as in Ref.~\cite{DeAmicis:2024not} (radial step $\Delta\rho=0.015$) and refer to this work for a detailed investigation of the code convergence and error budget quantification.
We will present analytical predictions for the first component of the time-domain Green's function, the one observed right before the quasinormal ringing, commonly denoted \textit{prompt response}.
To disentangle these different components, it is standard to study the problem in the complex frequency domain, as first proposed and implemented by Leaver~\cite{Leaver:1986gd}.
To move into the complex-frequency domain, we define the Fourier transform and its anti-transform as
\begin{equation}
\begin{split}
       & \tilde{G}_{\ell m}(\omega;r_*,r_*') = \int_{t'}^{\infty}dt \  e^{i\omega(t-t')}G_{\ell m}(t-t';r_*,r_*')\, , \\
       &G_{\ell m}(t-t';r_*,r_*')  = \frac{1}{2\pi}\int_{-\infty}^{\infty}d\omega \, e^{-i\omega(t-t')}\tilde{G}_{\ell m}(\omega;r_*,r_*')\, ,
\end{split}
\label{eq:transform_t}
\end{equation}
then perform an analytic continuation of $\omega$ to complex values.
The frequency domain GF, $\tilde{G}_{\ell m}(\omega;r_*,r_*')$, can be written as:
\begin{equation}
    \tilde{G}_{\ell}(\omega;r_*,r_*')=\theta(r_*-r_*')\frac{u^{\rm in}_{\ell}(\omega, r_*')u^{\rm up}_{\ell}(\omega, r_*)}{W(\omega)} + \theta(r_*'-r_*)\frac{u^{\rm up}_{\ell}(\omega, r_*')u^{\rm in}_{\ell}(\omega, r_*)}{W(\omega)}
    \label{eq:tildeGF_ansatz}
\end{equation}
where 
$u^{\rm in}_{\ell}(\omega, r_*)$ and $u^{\rm up}_{\ell}(\omega, r_*)$ are two independent solutions of the homogeneous RWZ equation with boundary conditions
\begin{equation}
u_{\ell m}^{\rm in}(\omega,r_*) =
\begin{cases}
& e^{-i\omega r_*} \ \ , \ \ r_*\rightarrow-\infty\\
& A_{\rm in}(\omega) e^{-i\omega r_*} + A_{\rm out}(\omega) e^{i\omega r_*} \ \ , \ \ r_*\rightarrow\infty
\end{cases} \, , 
\label{eq:u_in_asympt_expr}
\end{equation}
\begin{equation}
u_{\ell m}^{\rm up}(\omega,r_*)  =
\begin{cases}
& B_{\rm in}(\omega) e^{-i\omega r_*} + B_{\rm out}(\omega) e^{i\omega r_*} \ \ , \ \ r_*\rightarrow-\infty\\
& e^{i\omega r_*} \ \ , \ \ r_*\rightarrow\infty \, .
\end{cases}
\label{eq:u_out_asympt_expr}
\end{equation}
and $W(\omega)\equiv u^{\rm up} \partial_{r_*}u^{\rm in}-u^{\rm in} \partial_{r_*}u^{\rm up} $ is their Wronskian.
Since we will focus on an observer at $\mathcal{I}^+$, such that $r_*>r_*'$, we will from now on neglect the second term on the right-hand side in equation~\eqref{eq:tildeGF_ansatz}.

Following Leaver~\cite{Leaver:1986gd}, it is useful to introduce a third homogeneous solution to Eq.~\eqref{eq:RWZ_equation}, defined as
\begin{equation}
u_{\ell m}^{\infty-}(\omega,r_*)  =
\begin{cases}
& C_{\rm in}(\omega) e^{-i\omega r_*} + C_{\rm out}(\omega) e^{i\omega r_*} \ \ , \ \ r_*\rightarrow-\infty\\
& e^{-i\omega r_*} \ \ , \ \ r_*\rightarrow\infty \, .
\end{cases}
\label{eq:u_inf-_asympt_expr}
\end{equation}
The solutions $u^{\rm in},\,u^{\rm up}$ and $u^{\infty-}$ are not all independent, they are only pairwise independent, and each of them can be written as a combination of the other two.
Comparing the asymptotic behaviors in Eqs.~\eqref{eq:u_in_asympt_expr},~\eqref{eq:u_out_asympt_expr} and~\eqref{eq:u_inf-_asympt_expr} in the large distance limit $r_*\rightarrow+\infty$, we see that
\begin{equation}
u^{\rm in}_{\ell m}(\omega,r_*)=A_{\rm in}(\omega)u^{\infty -}_{\ell m}(\omega,r_*)+A_{\rm out}(\omega)u^{\rm up}_{\ell m}(\omega,r_*) \, .
\label{eq:u_in_vs_u_out_u_inft-}
\end{equation}
Substituting the above expansion in Eq.~\eqref{eq:tildeGF_ansatz}, the time domain GF can be computed as
\begin{equation}
G_{\ell m,\, r_*> r_*'}(t-t';r_*,r_*')=
G^{(1)}_{\ell m,\, r_*> r_*'}(t-t';r_*,r_*')
+ G^{(2)}_{\ell m,\, r_*> r_*'}(t-t';r_*,r_*')\, ,
\label{eq:GF_time_anti_transform_II_generic}
\end{equation}
where we have defined the following quantities
\begin{equation}
G^{(1)}_{\ell m,\, r_*> r_*'}(t-t';r_*,r_*')\equiv\frac{1}{2\pi} \int_{-\infty}^{\infty}d\omega \, \tilde{G}^{(1)}(\omega;r_*,r_*')\equiv\frac{i}{2\pi}\int_{-\infty}^{\infty}d\omega \, \frac{e^{-i\omega(t-r_*-t')}}{2\omega}u^{\infty-}(\omega,r_*') \, ,
\label{eq:G1_def}
\end{equation}
\begin{equation}
G^{(2)}_{\ell m,\, r_*> r_*'}(t-t';r_*,r_*')\equiv\frac{1}{2\pi} \int_{-\infty}^{\infty}d\omega \, \tilde{G}^{(2)}(\omega;r_*,r_*')\equiv\frac{i}{2\pi} \int_{-\infty}^{\infty}d\omega \, \frac{e^{-i\omega(t-r_*-t')}}{2\omega}\frac{A_{\rm up}(\omega)}{A_{\rm in}(\omega)}u^{\rm up}(\omega,r_*') \, .
\label{eq:G2_def}
\end{equation}
The term $G^{(2)}$ yields the QNMs response and Price's law, reaching $\mathcal{I}^+$ only after a retarded time~\cite{DeAmicis:2025xuh}
\begin{equation}
    t-r_*\geq t'+r_*'-4\log \left(\frac{r_*'-2}{r_*'}\right) \, .
\end{equation}
When the initial impulse (located at $(t',r_*')$) is outside the light-ring, the above causality condition becomes a scattering prescription from near the potential barrier peak. 
This propagation happens well inside the light-cone, while numerical experiments of the Green's function, see e.g. Fig.~\ref{fig:test_GFlat}, show that the prompt travels on and inside the curved light-cone, and is present until when the ringdown starts. We then argue that the prompt response originates from $G^{(1)}$ in Eq.~\eqref{eq:GF_time_anti_transform_II_generic}.
%

\section{Flat spacetime limit}   %
\label{sec:flat_spacetime} %
%

%
To gain some intuition of the nature of the prompt response and its relation with $G^{(1)}$ of Eq.~\eqref{eq:GF_time_anti_transform_II_generic}, we first restrict our attention to impulses localized very far away from the BH, where the spacetime can be approximated as flat. 
In the flat spacetime limit, the RWZ equations are identical and reduce to the following
\begin{equation}
\psi''(z)+\left(1-\frac{\ell (\ell+1)}{z^2}\right)\psi(z)=0 \, ,
\label{eq:RWZ_Flat}
\end{equation}
where we have introduced a new coordinate $z\equiv r\omega$.
The equation above corresponds to a Coulomb wave equation with $\eta=0$, hence its solutions are Coulomb wave functions with $\eta=0$ (see Appendix~\ref{app:Coulomb}). Following their asymptotic prescriptions, we can write the solutions $u^{\rm in,up,\infty -}$ as \footnote{Note that in the flat spacetime approximation, there is no BH so an ingoing solution at the horizon has no physical meaning. It is natural, in this limit, to assume a regular purely ingoing mode at the origin $r=0$ instead.}
\begin{equation}
\begin{split}
&u^{\infty-}_{\ell}(\omega,r_*)=H^{-}_{\ell}(0,z) \mathcal{N}^{\infty-}_{\ell}(\omega)\, , \\
&u^{\rm up}_{\ell}(\omega,r_*)=H^{+}_{\ell}(0,z) \mathcal{N}^{\rm up}_{\ell}(\omega)\ \, ,\\
&u_{\ell}^{\rm in}(\omega,r_*)= \frac{2^{\ell }|\Gamma(\ell+1)|}{\Gamma(2\ell+2)}z^{\ell+1}e^{-i\omega r}\mathrm{M}(\ell+1,2\ell+2,2iz) \mathcal{N}^{\rm in}_{\ell}(\omega)\, ,
\end{split}
\end{equation}
where $\mathcal{N}_{\ell}(\omega)$ are normalization factors, $\mathrm{M}$ is the Kummer's function and $H^{\pm}$ are superposition of the two Coulomb wave functions $F_{\ell},\, G_{\ell}$. We refer to Appendix~\ref{app:Coulomb} for more details.
First of all, we note that it holds
\begin{equation}
    H^{\pm}_{\ell }(0,z)=e^{\pm iz\mp i\pi \ell/2}(\mp 2iz)^{\ell +1}U(\ell+1,2\ell+2;\mp2iz)\, ,
\end{equation}
where 
\begin{equation}
    U(\ell+1,2\ell+2;\mp2iz)=(\mp2iz)^{-\ell-1}\sum_{s=0}^{\ell}\binom{\ell }{s}(\ell+1)_s(\mp 2iz)^{-s} \, ,
\end{equation}
is the Tricomi confluent hypergeometric function~\cite{abramowitz1968handbook,NIST:DLMF}.
Hence, the normalization factors for the solutions $u^{\infty-}_{\ell},\, u^{\rm up}_{\ell}$ are $\mathcal{N}^{\infty-}_{\ell}(\omega)=e^{-i\pi\ell/2},\, \mathcal{N}^{\rm up}_{\ell}(\omega)=e^{i\pi\ell/2}$, respectively. 

In the limit $r\rightarrow 0$ with generic $\omega<\infty$, it holds $z\ll1$ and $M(\ell+1,2\ell+2;2iz)\approx1+\mathcal{O}(z)$. 
Then, the normalization of $u^{\rm in}$ is
\begin{equation}
    \mathcal{N}^{\rm in}_{\ell}(\omega)=\left[\frac{2^{\ell+2}\Gamma(\ell+1)}{\Gamma(2\ell+2)}\omega^{\ell+1}\right]^{-1}\, .
\end{equation}

\begin{figure*}[h!]
\includegraphics[width=0.99\textwidth]{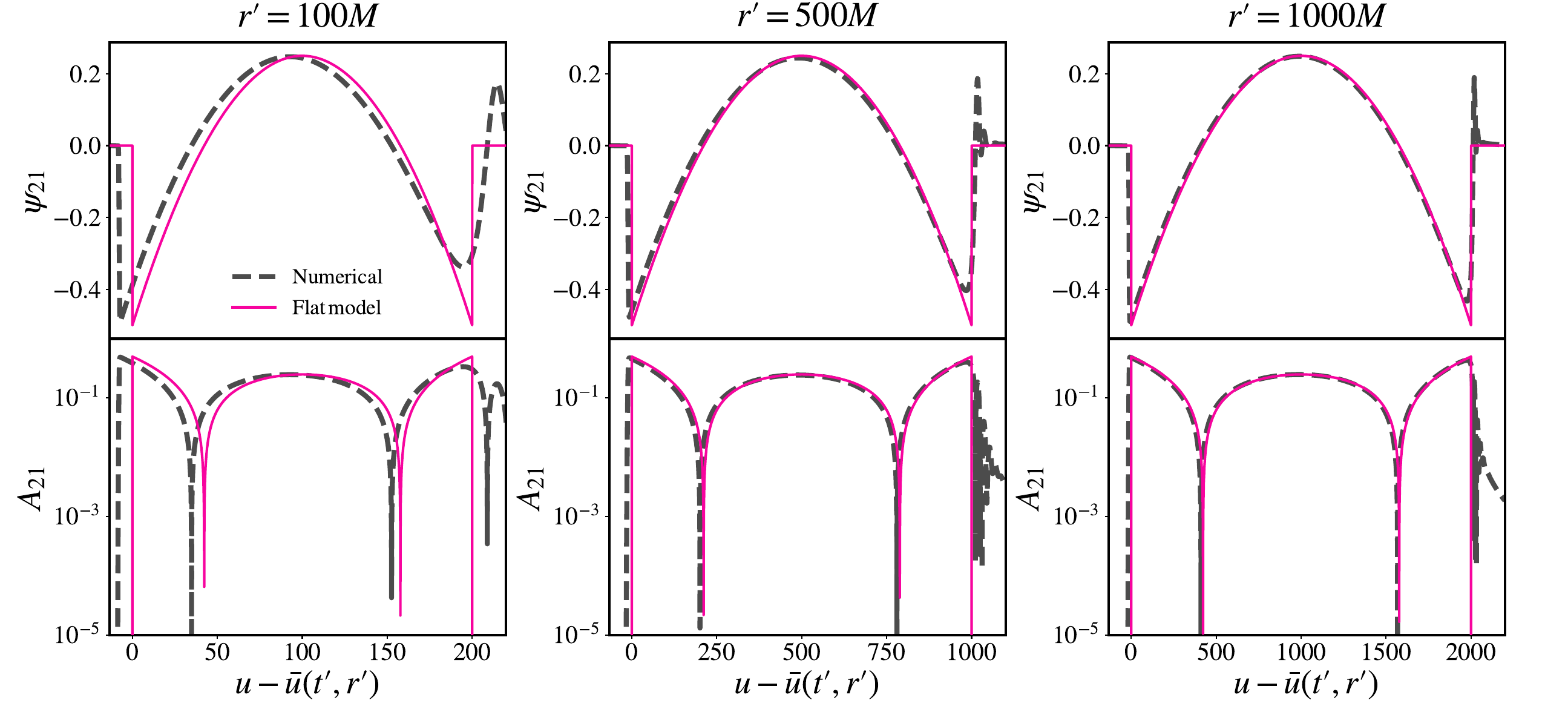}
\caption{Real component ('$+$' polarization) (\textbf{top}) and amplitude (\textbf{bottom}) of the Green's function $\ell=2,\,m=1$ multipole as observed at $\mathcal{I}^+$ vs the retarded time of the observer $u$. This quantity has been rescaled with the retarded time at which the first signal travelling on the flat light cone would arrive.
Results shown for different location of the initial impulse $r'$, as displayed in the labels.
In dashed grey, results of numerical integrations obtained through the \textsc{RWZHyp} code~\cite{Bernuzzi:2010ty,Bernuzzi:2011aj}, as described in Sec.~\ref{sec:background}.
In bright purple, the prediction obtained in the flat spacetime limit.
\label{fig:test_GFlat}}
\end{figure*}

In the case $\ell=2$, it holds:
\begin{equation}
U(3,6;\mp2iz)=(\mp2iz)^{-3}\left(1\pm\frac{3i}{z}-\frac{3}{z^2}\right)\, .
\end{equation}
Hence:
\begin{equation}
    \begin{split}
        & u^{\infty-}_{\ell=2}(\omega,r)=e^{-i\omega r}\left(-\frac{3}{\omega^2 r^2}-\frac{3i}{\omega r}+1\right)\, , \\
        & u^{\rm up}_{\ell=2}(\omega,r)=e^{i\omega r}\left(-\frac{3}{\omega^2 r^2}+\frac{3i}{\omega r}+1\right)\, .
    \end{split}
    \label{eq:uinupinf-_ell_2_om_zero}
\end{equation}
Furthermore, for $r \rightarrow \infty$, it holds:
\begin{equation}
    u^{\rm in}_{\ell=2}(\omega,r)=\frac{i \ \mathcal{N}^{\rm in}_{\ell=2}(\omega)}{2}\left[e^{-i\omega r}\left(\frac{3}{\omega^2 r^2}+\frac{3i}{\omega r}-1\right)+e^{i\omega r}\left(-\frac{3}{\omega^2 r^2}+\frac{3i}{\omega r}+1\right) \right]\, .
\end{equation}
Comparing the above expressions with Eq.~\eqref{eq:u_in_vs_u_out_u_inft-}, we find
\begin{equation}
A_{\ell=2,\,\rm in}(\omega)= -\frac{i}{2} \mathcal{N}_{\ell=2}^{\rm in}(\omega )\  , \, A_{\ell=2,\,\rm up}(\omega)=\frac{i}{2} \mathcal{N}_{\ell=2}^{\rm in}(\omega )\, .
\label{eq:AinAout_ell2_om_zero}
\end{equation}
Substituting Eqs.~\eqref{eq:uinupinf-_ell_2_om_zero} and~\eqref{eq:AinAout_ell2_om_zero} into Eq.~\eqref{eq:GF_time_anti_transform_II_generic} (rewritten for the flat spacetime limit) we find 
\begin{equation}
\begin{split}
G_{\ell=2\, m,\, r> r'}(t-t';r,r')=&\frac{1}{2\pi}\int_{-\infty}^{\infty}d\omega \, e^{-i\omega(\bar{u}(t,r)-\bar{u}(t',r'))}\left(\frac{i}{2\omega}+\frac{3}{2\omega^2 r'}-\frac{3i}{2\omega^3 r'^2}\right)+\\
&\frac{1}{2\pi}\int_{-\infty}^{\infty}d\omega \, e^{-i\omega(\bar{u}(t,r)-\bar{v}(t',r'))}\left(-\frac{i}{2\omega}+\frac{3}{2\omega^2 r'}+\frac{3i}{2\omega^3 r'^2}\right)\, ,
\end{split}
\label{eq:GF_time_anti_transform_II_om0}
\end{equation}
where we have introduced the flat retarded coordinates as $\bar{u}(t,r)\equiv t-r,\,\bar{v}(t,r)\equiv t+r$.
We denote the integral on the first line of Eq.~\eqref{eq:GF_time_anti_transform_II_om0} as $G^{(1)}$, and the one on the second line as $G^{(2)}$.
Both integrands are singular for $\omega=0$, to compute the integrals we perform an analytical continuation in the complex-$\omega$ plane.
$G^{(1),(2)}$ are then computed along an axis which is parallel to the real-$\omega$ one, shifted by a small $\epsilon$ quantity on the positive imaginary-$\omega$ axis, i.e. $\Gamma = \{\omega \in \mathbb{R} + i\epsilon\}$ with 
$\epsilon \ll 1$.
The small positive imaginary part implements the retarded 
Green's--function prescription.
We evaluate $G^{(1),(2)}$ by means of the residue theorem, choosing a prescription to close the complex contour.
For $\bar{u}(t,r)<\bar{u}(t',r')$ ($\bar{u}(t,r)<\bar{v}(t',r')$), the contour for $G^{(1)}$ ($G^{(2)}$) is closed with a semi circle of radius $|\omega|\rightarrow\infty$ in the upper 
half plane. The contour does not include any pole and the integrand vanishes for $|\omega|\rightarrow\infty$. Then, due to Jordan's lemma, the integral along the arc (hence the integrand along the real axis) vanishes. 
For $\bar{u}(t,r)>\bar{u}(t',r')$ ($\bar{u}(t,r)>\bar{v}(t',r')$) 
the contour for $G^{(1)}$ ($G^{(2)}$) is closed in the lower half plane. The integral along the infinite-radius arc vanishes (again due to Jordan's lemma) and the only contribution to the integral along the real axis comes from the pole in $\omega=0$
\begin{equation}
\begin{split}
    & G_{\ell=2\, m ,r>r'}^{(1)}=\theta(\bar{u}(t,r)-\bar{u}(t',r'))\cdot \left[\frac{3}{4(r')^2}(\bar{u}(t,r)-\bar{u}(t',r'))^2 -\frac{3}{2r'}(\bar{u}(t,r)-\bar{u}(t',r'))+\frac{1}{2} \right]\\
    &G_{\ell=2\, m ,r>r'}^{(2)}=\theta(\bar{u}(t,r)-\bar{v}(t',r'))\cdot \left[-\frac{3}{4 (r')^2}(\bar{u}(t,r)-\bar{v}(t',r'))^2 -\frac{3}{2r'}(\bar{u}(t,r)-\bar{v}(t',r'))-\frac{1}{2} \right]
\end{split}
\label{eq:G1_G2_results_flat}
\end{equation}
These contributions (when both are present) perfectly cancel. However, there exists a regime for which $\bar{u}(t,r)>\bar{u}(t',r')$ but $\bar{u}(t,r)<\bar{v}(t',r')$ 
during which
\begin{equation}
    G_{\ell=2\, m,r>r'}^{\rm Flat}(t-t';r,r')=\theta\left(\bar{u}(t,r)-\bar{u}(t',r')\right)\theta\left(\bar{v}(t',r')-\bar{u}(t,r)\right)\cdot \left[-\frac{1}{4} + \frac{3}{4(r')^2}\left(\bar{u}(t,r)-t'\right)^2 \right]
    \label{eq:GFlat}
\end{equation}
Eq.~\eqref{eq:RWZ_Flat}, can be obtained as a particular limit of the RWZ problem for $\omega M\ll1$ and $\omega r\gg1$, as a zero-order approximation where we neglect all terms $\mathcal{O}(M)$.
This implies that we can obtain a prediction valid only for a time $M\ll t\ll r$: if $r$ is sufficiently far away, it is possible to find a time interval where both conditions are satisfied, and our approximation can be used. As $r$ gets closer to the BH, the approximation starts to fail.
We have tested Eq.~\eqref{eq:GFlat} against numerical evolutions obtained by solving Eq.~\eqref{eq:GF_definition}. The results are shown in Fig.~\ref{fig:test_GFlat}.
As expected, the predicted parabolic-like behavior in Eq.~\eqref{eq:GFlat} well reproduces the prompt response for an initial location of the compact source $\sim\mathcal{O}(10^3)$ or larger, and starts failing on shorter scales.
In particular, Fig.~\ref{fig:test_GFlat} shows a growing time shift between the numerical waveform and the flat spacetime prediction, while the overall morphology has minor modifications.
The flat-spacetime propagator components in a spin-2 weight spherical harmonic decomposition, i.e. Eq.~\eqref{eq:GFlat}, was already computed in Ref.~\cite{Bonga:2018zlx} with a purely time-domain computation.
Instead, we carry our analysis in the frequency domain, and anti-transform back to the time domain only at the end.
Our results are consistent with those of Ref.~\cite{Bonga:2018zlx} and allow for the first time to associate the flat spacetime propagator with a pole in $\omega=0$ in the complex frequency domain. 
In the past literature, the prompt response of a Schwarzchild BH has always been (intuitively, without proof) associated with high-frequency arcs in the complex-frequency domain. 
Since Eq.~\eqref{eq:GFlat} reproduces the prompt to sources localized at very large distances from the BH, 
our computation clearly shows that the prompt is rather a small-frequency response and the contribution of the arcs, in this limit, is zero.
%
%
%
%

%

\section{Schwarzschild: odd sector}   %
\label{sec:Schwarzschild} %
%

%
We now wish to compute the prompt response of a Schwarzschild black hole, for compact sources driving the perturbations localized far from the BH.
To achieve this, we employ solutions of the RWZ equation in terms of superposition of Coulomb wave functions, originally proposed by Leaver~\cite{Leaver:1986vnb} to study the small-frequency response (the one activated at large distances).
Given the RWZ problem in Eq.~\eqref{eq:RWZ_equation}, we introduce the coordinate $z\equiv \omega r$ and define the field variable $y(z(r))\equiv A^{2i\omega M}(r) \, \psi(r)$, satisfying the equation
\begin{equation}
z(z-2\omega M)\left[y_{,zz}(z)+\left(1-\frac{2\eta}{z}\right)y(z)\right]+\omega M\, C_1(\omega)y_{,z}(z)+\left(C_2(\omega)+\frac{\omega M\, C_3(\omega)}{z}\right)y(z) = 0
\label{eq:RWZ_for_y_field_variable}
\end{equation}
where we have used the notation $y_{,z}\equiv\partial_zy$ to denote the derivative with respect to $z$.
The coefficients $C_{1,2,3},\,\eta$ are functions of $\omega$, defined as
\begin{equation}
C_1\equiv 2 - 8 i \omega M \ \ , \ \ \ \ C_2\equiv -\ell(\ell+1) + 12 \omega^2 M^2 \ \ , \ \ \ \ C_3\equiv 6 + 8 i \omega M  + 8 \omega^2 M^2\ \ , \ \ \ \ \eta=-2\omega M \, .
\end{equation}
Following Ref.~\cite{Leaver:1986vnb}, solutions to the RWZ equation can be written as
\begin{equation}
\psi(r(z))=\mathcal{N}(\omega) \, \left(1-\frac{2\omega M}{z}\right)^{-2i\omega M}\, \sum_{L=-\infty}^{\infty}a_{L}(\omega) \, \mathcal{U}_{L+\nu}(\eta,z) \, ,
\label{eq:psi_as_coulomb_superposition}
\end{equation}
where $\mathcal{U}_{L+\nu}(\eta,z)$ is a superposition of the Coulomb wave functions $F_{L+\nu}(\eta,z),\,G_{L+\nu}(\eta,z)$ as defined in Chapter 14 of Ref.~\cite{abramowitz1968handbook}, $\nu$ is an $\omega$-dependent parameter and the coefficients $a_L$ are solutions to the three-terms recursive relation
\begin{equation}
\alpha_L a_{L+1}+\beta_La_L+\gamma_La_{L-1}=0\, ,
\label{eq:recursive_relation}
\end{equation}
with boundary condition $a_0=1$.
We refer to~\cite{Leaver:1986vnb} for an explicit expression of the coefficients $\alpha_L,\beta_L,\gamma_L$ that appear in the equation above.
We perform a Taylor expansion of the recursive equation above in the limit $\omega M\rightarrow0$, and solve iteratively up to order $\mathcal{O}[(\omega M)^4]$, following a strategy similar to Refs.~\cite{Mano:1996mf,Casals:2015nja}.
The results for the coefficients $\nu,\, a_L$ up to order $\mathcal{O}[(\omega M)^1]$ can be written as follows\footnote{In this formalism, the "renormalized" angular momentum satisfies the symmetry 
$\nu \leftrightarrow -\nu-1$ (both roots solve the defining equation). 
At zeroth PM order, we choose $\nu=\ell$ rather than $\nu=-\ell-1$, so that the solution 
reduces smoothly to the flat–spacetime limit.}
\begin{equation}
\begin{split}
&a_1=\frac{(-3 + 2 \ell + \ell^2) \,\omega M}{(1 + \ell) (1 + 2 \ell)}  \, ,\\
&a_{-1}= \frac{(4 - \ell^2) \, \omega M}{\ell + 2 \ell^2} \, ,\\
&\nu = \ell+\mathcal{O}[(\omega M)^2] \, .
\end{split}
\end{equation}
It holds $a_L\sim\mathcal{O}[(\omega M)^{|L|}]$, so at first order, all the other coefficients $a_L,\, |L|>1$ vanish.
The explicit expressions of the Taylor-series coefficients $\nu,\,a_L$ up to order $\mathcal{O}[(\omega M)^4]$ can be found in Appendix~\ref{app:aL_coeffs}.\\
We can now write the solutions to the RWZ equations, $u^{\rm in,up,\infty-}$ as defined in Eqs.~\eqref{eq:u_in_asympt_expr},~\eqref{eq:u_out_asympt_expr} and~\eqref{eq:u_inf-_asympt_expr}, in an expansion in the limit of $\omega M\rightarrow 0$, for generic locations of the source $r'$.
The $u^{\rm in}_{\ell}$ solution is written in terms of the Coulomb wave functions $F_{L+\nu}(\eta,z)$~\cite{Leaver:1986vnb}, which in turn can be expressed through the Kummer's function $\mathrm{M}(a,b,z)$ (see Appendix ~\ref{app:Coulomb}), so that
\begin{equation}
\begin{split}
u^{\rm in}_{\ell}(\omega,r)&=\,\mathcal{N}^{\rm in}(\omega) \, \left(1-\frac{2\omega M}{z}\right)^{-2i\omega M}\, \sum_{L=-\infty}^{\infty}a_{L}(\omega) \, F_{L+\nu}(\eta,z)=\mathcal{N}^{\rm in}(\omega) \, e^{-iz-2i \omega M\log\left(1-\frac{2\omega M}{z}\right)}\times\\
& \, \sum_{L=-\infty}^{\infty}a_{L}(\omega) \, \frac{\left[\Gamma(L+\nu+1+i\eta)\Gamma(L+\nu+1-i\eta)\right]^{1/2}}{2e^{\pi\eta/2}\,\Gamma(2L+2\nu+2)} (2z)^{L+\nu+1}\, \mathrm{M}(L+\nu+1-i\eta,2L+2\nu+2,2iz)\, .
\end{split}
\label{eq:uIN_coulomb_superposition}
\end{equation}
In the limit $r\rightarrow+\infty$, i.e. for $z\rightarrow+\infty$, the Kummer's function admits the asymptotic expansion
\begin{equation}
\begin{split}
\mathrm{M}(L+\nu+1-i\eta,2L+2\nu+2,2iz)&=(2iz)^{-L-\nu-1+i\eta}\,e^{i\pi(L+\nu+1-i\eta)}\frac{\Gamma{(2L+2\nu+2)}}{\Gamma{(L+\nu+1+i\eta)}}\times\\
&\sum_{n=0}^{\infty}\frac{1}{n!}(L+\nu+1-i\eta)_n(-L-\nu+i\eta)_n(-2iz)^{-n}+e^{2iz}\,(2iz)^{-L-\nu-1-i\eta}\times\\
&\frac{\Gamma{(2L+2\nu+2)}}{\Gamma{(L+\nu+1+i\eta)}}\sum_{n=0}^{\infty}\frac{1}{n!}(L+\nu+1+i\eta)_n(-L-\nu+i\eta)_n(2iz)^{-n}\, ,
\end{split}
\end{equation}
restoring $u^{\rm in}_{\ell}$ expression as superposition of $u^{\rm up}_{\ell}$ and $u^{\infty -}_{\ell}$, as shown in Eq.~\eqref{eq:u_in_vs_u_out_u_inft-}.
In the limit $r\rightarrow 2M$, the solution is dominated by the factor $A^{-2i\omega M}(r)$: it behaves as a purely ingoing wave at the horizon.
The normalization $\mathcal{N}^{\rm in}(\omega)$ ensures that this wave is also unitary $u^{\rm in}_{\ell}(\omega,r_*\rightarrow-\infty)\rightarrow e^{-i\omega r_*}$.
The solutions $u^{\rm up,\infty-}_{\ell}$ can be written as  superposition of $H^{\pm}_{L+\nu}(\eta,z)\equiv G_{L+\nu}(\eta,z)\pm i F_{L+\nu}(\eta,z)$, respectively~\cite{Leaver:1986vnb}
\begin{equation}
u^{\rm up}_{\ell}(\omega,r)=\mathcal{N}^{\rm up}(\omega) \, \left(1-\frac{2\omega M}{z}\right)^{-2i\omega M}\, \sum_{L=-\infty}^{\infty}a_{L}(\omega) \, H^+_{L+\nu}(\eta,z) \, ,
\label{eq:u_up_coulomb_superposition}
\end{equation}
\begin{equation}
u^{\infty -}_{\ell}(\omega,r)=\mathcal{N}^{\infty-}(\omega) \, \left(1-\frac{2\omega M}{z}\right)^{-2i\omega M}\, \sum_{L=-\infty}^{\infty}a_{L}(\omega) \, H^-_{L+\nu}(\eta,z) \, ,
\label{eq:u_infm_coulomb_superposition}
\end{equation}
In the limit $r\rightarrow\infty$, $u^{\rm up}_{\ell},\,u^{\infty-}_{\ell}$ behave, respectively, as a pure in- ($+$),  out- ($-$) going wave, since the Coulomb wave functions satisfy~\cite{abramowitz1968handbook,Leaver:1986vnb}
\begin{equation}
\lim_{z\rightarrow\infty} H^{\pm}_{L+\nu}(\eta,z)\rightarrow\exp\left\lbrace\pm i\left(z-\eta\log 2z - (L+\nu)\pi/2\right)\right\rbrace  \, \left[\frac{\Gamma(L+\nu+1+i\eta)}{\Gamma(L+\nu+1-i\eta)}\right]^{\pm 1/2}\, ,
\end{equation}
The normalization factors $\mathcal{N}^{\rm up}(\omega),\,\mathcal{N}^{\infty-}(\omega)$ are fixed such that the solutions $u^{\rm up}_{\ell},\,u^{\infty-}_{\ell}$ are unitary waves in this limit, as required in Eqs.~\eqref{eq:u_out_asympt_expr} and~\eqref{eq:u_inf-_asympt_expr}.
We show their generic expression and the Taylor expansion for small $\omega M$ with corrections up to $\mathcal{O}(\omega M)$
\begin{center}
\begin{equation}
\begin{split}
\mathcal{N}^{\infty-}(\omega)=&2^{4i\omega M}e^{-i\nu\pi+\omega M\pi}(i\omega M)^{2i\omega M+\nu}(\omega M)^{-\nu} \, \left\lbrace\sum_{L=-\infty}^{\infty}i^L\, a_L(\omega)\, \left[\frac{\Gamma(\nu+L+1+2i\omega M)}{\Gamma(\nu+L+1-2i\omega M)}\right]^{1/2}\right\rbrace^{-1}=\\
&=-1 - \frac{i }{3} \left[-10 + 6 \gamma - 3 i \pi + \log(4096) + 
    6 \log(i \omega M)\right]\,\omega M+\mathcal{O}[(\omega M)^2]\, ,
\end{split}
\label{eq:Ninf-_expr}
\end{equation}
\begin{equation}
\begin{split}
\mathcal{N}^{\rm up}(\omega)=&2^{-4i\omega M}e^{i\nu \pi+\omega M\pi}(-i \omega M)^{\nu-2i\omega M}(\omega M)^{-\nu} \, \left\lbrace\sum_{L=-\infty}^{\infty}i^{-L}\, a_L(\omega)\, \left[\frac{\Gamma(\nu+L+1-2i\omega M)}{\Gamma(\nu+L+1+2i\omega M)}\right]^{1/2}\right\rbrace^{-1}=\\
&=-1 + \frac{1}{3}\left[-10 + 6 \gamma + 3 i\pi + \log(4096) + 6 \log(-i \omega M)\right]\, \omega M + \mathcal{O}[(\omega M)^2]\, .
\end{split}
\label{eq:Nup_expr}
\end{equation}
\end{center}
These normalization terms contain multi-valued factors that in a small $\omega M$ expansion are decoupled from the single-valued behaviour, and appear starting from the first order as $\propto\log(\pm i\omega M)$ contributions.
To prepare for the computations in the next sections, we show that it is possible to express Eqs.~\eqref{eq:u_up_coulomb_superposition},~\eqref{eq:u_infm_coulomb_superposition} in terms of the Tricomi confluent hypergeometric function $U(a,b,z)$ (as defined in, e.g., Eq.~(13.1.3) of Ref.~\cite{abramowitz1968handbook}). This can be achieved through (see Sec.~IV of Ref.~\cite{Leaver:1986vnb})
\begin{equation}
H^{\pm}_{L+\nu}(\eta(\omega),z)= C_{L,\,\nu}^{\pm}(\omega)
\,\, (2z)^{L+\nu+1}e^{\pm i z}\, U(L+\nu+1\pm i \eta,2L+2\nu+2,\mp 2 i z)  \, ,
\label{eq:Hpm_vs_irregular_confluent_hypergeo}
\end{equation}
with 
\begin{equation}
C_{L,\,\nu}^{\pm}(\omega)\equiv \frac{(-1)^{L}}{e^{\pi\omega M}\, e^{\pm i \pi (\nu+1/2)}}\left[\frac{\Gamma(L+\nu+1-i2\omega M)}{\Gamma(L+\nu+1+i2\omega M)}\right]^{\pm 1/2} \,.
\end{equation}

\subsection{Computation of $G^{(1)}_{\ell m}$}
\label{subsec:G1}
We focus on computing $G^{(1)}_{\ell m,,r_*>r_*'}$, as defined in Eq.~\eqref{eq:G1_def}.
We rewrite $u^{\infty-}_{\ell}$ in terms of Kummer's functions $M(a,b,z)$, by exploiting, in Eq.~\eqref{eq:Hpm_vs_irregular_confluent_hypergeo}, the following transformation (see e.g. Eq.~(13.1.13) of Ref.~\cite{abramowitz1968handbook})
\begin{equation}
U(a,b,z)=\frac{\pi}{\sin(\pi b)}\left[\frac{\mathrm{M}(a,b,z)}{\Gamma(1+a-b)\Gamma(b)}-z^{1-b}\frac{\mathrm{M}(1+a-b,2-b,z)}{\Gamma(a)\Gamma(2-b)} \right] \, ,
\label{eq:Kummer_transformation_generic_def}
\end{equation}
valid for all $b$ that are not integers. 
In the case of $u^{\infty-}$, the above expression can be used if we consider $\nu$ to be a generic complex function of $\omega M$, rather than restrict purely to the zeroth or first order-truncated expansions in $\omega M$ \footnote{As stated above, the representation of $U$ we adopt can be used when $\nu$ is not an integer, and works in the $O(\omega M)$ limit assuming $\nu=l+\nu_2+...$, with $\nu_2\sim\mathcal{O} (\omega M)^2$, if one computes the $\nu\rightarrow \ell$ limit only after expanding in $\omega M$. An alternative representation valid when $b$ is an integer and therefore applicable up to $O(\omega M)$ in our computation can be found in  Eq.~(13.2.9) of Ref.~\cite{NIST:DLMF}. The results with the two procedures coincide, at first order.}. Then, we have
\begin{equation}
\begin{split}
U(L+\nu+1\pm i\eta,2L+2\nu+2,\mp2iz)&=\frac{\pi}{\sin[\pi(2L+2\nu+2)]}\left\lbrace\frac{\mathrm{M}(L+\nu+1\pm i\eta,2L+2\nu+2,\mp2iz)}{\Gamma(\pm i\eta -L-\nu)\Gamma(2L+2\nu+2)}-\right.\\
&\left.(\mp 2iz)^{-2L-2\nu-1}\frac{\mathrm{M}(\mp i\eta-L-\nu,-2L-2\nu,\mp2iz)}{\Gamma{(L+\nu+1\pm i\eta)}\Gamma{(-2L-2\nu)}}\right\rbrace\, .
\end{split}
\label{eq:Kummer_transformation_uinfm_uup}
\end{equation}
The first term on the right-hand side of Eqs.~\eqref{eq:Kummer_transformation_generic_def} and~\eqref{eq:Kummer_transformation_uinfm_uup} is single valued, while the second term contains a multi valued contribution.
In this work, we aim to provide a Post-Minkowskian expression of the prompt: a result that can be applied to compact sources localized far from the BH.
Such sources overlap mostly with the low frequency part of the GF, while influence from its high-frequency components is negligible.
For this reason, we expand the GF in the frequency-domain to the limit $\omega M\ll1$, then we compute the GF in the time-domain by integrating this expression into the entire complex domain $\omega M$.
Since high-frequency components of the GF are not greatly excited by compact sources localized far away from the BH, we expect the result to yield a good approximation of the signal emitted by such sources: the PM small $M/r\ll 1$ expansion justifies the $\omega M\ll1 $ one.
This PM approach allows us to divide the behavior near $\omega M=0$ into a multi-valued component and a single-valued one, both contributing to the time-domain GF. We show this below. \\
We insert Eqs.~\eqref{eq:u_infm_coulomb_superposition},~\eqref{eq:Hpm_vs_irregular_confluent_hypergeo} and~\eqref{eq:Kummer_transformation_uinfm_uup} into $\tilde{G}^{(1)}(\omega;r_*,r_*')$, Eq.~\eqref{eq:G1_def} 
\begin{equation}
\begin{split}
\tilde{G}^{(1)}&(\omega;r_*,r_*')=-M e^{-i\omega(t-r_*-t'-r')}A^{-2i\omega M}(r) \,\times \left\lbrace\right.
\\
& \left.\sum_{L=-\infty}^{\infty} a_L(\omega)\,(\omega M)^{L+1}\, \left[(i\omega M)^{-1+2i\omega M+\nu}\right]\left(\frac{r'}{M}\right)^{L+\nu+1}\mathcal{A}^-_{L,\nu}(\omega)\, \mathrm{M}(L+\nu+1+2i\omega M,2L+2\nu+2,2i\omega r') + \right.\\
&\left.+ \sum_{L=-\infty}^{\infty} a_L(\omega)\,(\omega M)^{-L} \, \left[(i\omega M)^{-2+2i\omega M-2L-\nu}(\omega M)^{2L+1}\right]\left(\frac{r'}{M}\right)^{-L-\nu}\, \mathcal{B}^-_{L,\nu}(\omega)\, \mathrm{M}(-2i\omega M-L-\nu,-2L-2\nu,2i\omega r')\right\rbrace\, ,
\end{split}
\label{eq:tildeG1_vs_kummer_fun}
\end{equation}
where we defined the coefficients
\begin{equation}\label{eq:AminusBminus}
\begin{split}
&\mathcal{A}^-_{L,\nu}(\omega)\equiv \tilde{N}^{\infty-}(\omega)\, C_{L,\nu}^-(\omega)\, 2^{L+\nu}\,  \frac{\pi}{\sin\left[2\pi(L+\nu+1)\right]}\,\frac{1}{\Gamma(2i\omega M-L-\nu)\Gamma(2L+2\nu+2)}\, , \\
&\mathcal{B}^-_{L,\nu}(\omega)\equiv \tilde{N}^{\infty-}(\omega)\, C_{L,\nu}^-(\omega)\, 2^{-1-L-\nu}\,  \frac{\pi}{\sin\left[2\pi(L+\nu+1)\right]}\,\frac{1}{\Gamma(L+\nu+1+2i\omega M)\Gamma(-2L-2\nu)}
\, ,
\end{split}
\end{equation}
and the single-valued component $\tilde{N}^{\infty-}$ of the normalization $\mathcal{N}^{\infty-}$ in Eqs.~\eqref{eq:Ninf-_expr}, factoring out from this expression the multi-valued ones, as
\begin{equation}
    \mathcal{N}^{\infty-}(\omega)\equiv (i\omega M)^{2i\omega M+\nu}(\omega M)^{-\nu}\tilde{N}^{\infty-}(\omega)\, .
\end{equation}
We discuss separately the complex $\omega M$-plane features of the second and third line in Eq.~\eqref{eq:tildeG1_vs_kummer_fun}, when expanded in a PM, $\omega M \rightarrow 0$, approximation.
In this limit, the coefficient $\mathcal{A}^-_{L,\nu}(\omega)$, as well as the Kummer's function $\mathrm{M}(L+\nu+1+2i\omega G,2L+2\nu+2,2i\omega r)$ are analytical and behave as\footnote{For finite values of $z$, the Kummer's function can be written as $\mathrm{M}(a,b,z)=1+\frac{a}{b}z+\frac{(a)_2}{(b)_2}\frac{z^2}{2}+...+ \frac{(a)_n}{(b)_n}\frac{z^n}{n!}+ ...$ with $(a)_n$ as the Pochhammer's symbol~\cite{abramowitz1968handbook}.} $\sim1+\mathcal{O}[(\omega M)]$.
The factor $a_L\, (\omega M)^L$ is analytical for all complex frequencies and behaves as $\sim\mathcal{O}[(\omega M)^{2L}]$ for $L>0$, and as $\sim 1+\mathcal{O}(\omega M)$ for $L<0$, since $a_L\sim(\omega M)^{|L|}$.
The term $\left[(i\omega M)^{2i\omega M+\nu}\right]$ contains multi-valued contributions, that can be isolated when working in a PM limit as complex logarithms $\sim (\omega M)^{\ell}[1+2i\omega M \log{(2i\omega M)}]$. 
Hence, working in a PM expansion, the second line of Eq.~\eqref{eq:tildeG1_vs_kummer_fun} contains a superposition of terms that are analytical on the whole complex plane, and multi-valued terms which introduce a branch cut along the negative imaginary axis, originating from the branch point at the origin.
The analytical terms give vanishing contribution to the time-domain GF due to Jordan's lemma: in fact, we always choose the contour so that the exponential $e^{-i\omega(t-r_*-t'+r')}$ is suppressed, rather than divergent.
The multi-valued terms, instead, give rise to a tail term, which is not the scope of the current work and will be explored elsewhere.
For this reason, in the following we will not compute the contribution of the terms on the second line of Eq.~\eqref{eq:tildeG1_vs_kummer_fun}, to the time-domain GF.

\begin{figure*}[t!]
\includegraphics[width=0.99\textwidth]{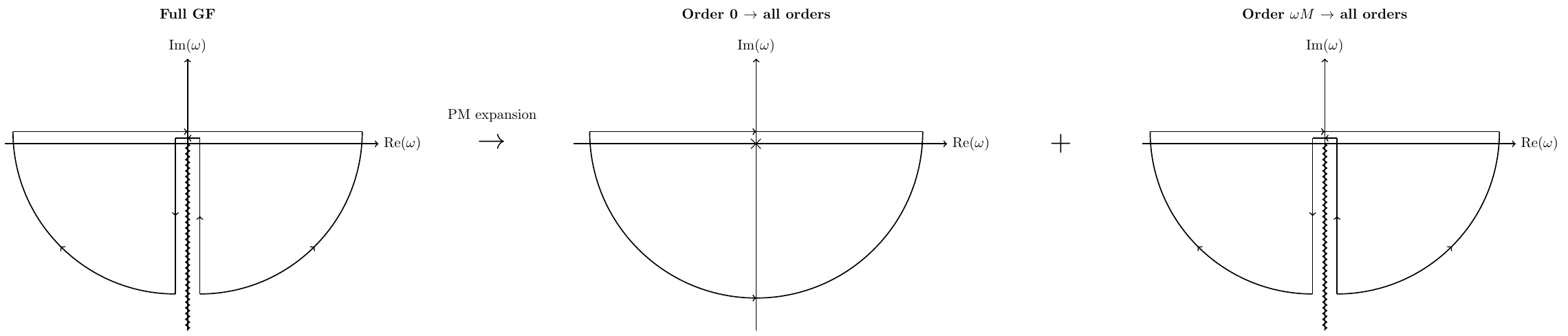}
\caption{Schematic representation of the non-analytical behavior of the Laplace-transformed Green's function component $\tilde{G}^{(1)}$, in the complex-frequency plane. 
In a PM expansion, this quantity is superposition of: a contribution which has integrable poles up to order $\ell+1$ in $\omega=0$, starting from the lowest PM order $\mathcal{O}[(\omega M)^0]$; a term with a branch cut along the negative imaginary axis, starting from the first PM order $\mathcal{O}(\omega M)$ 
\label{fig:contour_plot_prompt_branch}}
\end{figure*}

Now we move to the third line in Eq.~\eqref{eq:tildeG1_vs_kummer_fun}.
The term in squared brakets, when expanded in PM expansion, yields
\begin{equation}
(i\omega M)^{-2+2i\omega M-\nu-2L}(\omega M)^{2L+1}\approx -i^{-\ell-2L}\omega^{-1-\ell}+\mathcal{O}[(\omega M)^{-\ell-1}\log(i\omega M)] \, ,
\label{eq:poles_branch_cut_decoupled}
\end{equation}
recalling $\nu=\ell+\mathcal{O}[(\omega M)^2]$. The multi-valued terms $\propto \log(i\omega M)$ generate a branch cut on the negative imaginary $\omega$ axis, with branch point in $\omega=0$. These terms give rise to a tail behavior which we leave for future work.
The single-valued term, instead, gives rise to a series of poles, with order up to $-1-\ell$. These are the contributions we are interested in, to study the prompt. 
As schematically shown in Fig.~\ref{fig:contour_plot_prompt_branch}, the PM expansion allows to decouple the tail and prompt contribution near $\omega M\rightarrow 0$.
To show this in more detail, we substitute 
Eq.~\eqref{eq:poles_branch_cut_decoupled} into the third line of Eq.~\eqref{eq:tildeG1_vs_kummer_fun},
and focus on the single-valued contribution %
\begin{equation}
\begin{split}
\tilde{G}_{\ell m,\,r_*\gg r_*'}^{(1),\,\rm Prompt}(\omega;r_*,r_*')=&M\, i^{-\ell-2L}\,e^{-i\omega(t-r_*-t'+r')}A^{-2i\omega M}(r')\,\times\\
&\sum_{-2L<\ell} a_L(\omega)\,(\omega M)^{-L-\ell-1} \, \left(\frac{r'}{M}\right)^{-L-\nu}\, \mathcal{B}^-_{L,\nu}(\omega)\, \mathrm{M}(2i\omega M-L-\nu,-2L-2\nu,2i\omega r')\, ,
\end{split}
\label{eq:tildeG1_prompt}
\end{equation}
In the above, the coefficients $\mathcal{B}^-_{L,\nu}$ and the Kummer's functions are analytical, and behave as $1+\mathcal{O}(\omega M)$ in the limit $\omega M \rightarrow 0$.
The factor $a_L\, (\omega M)^{-L}$, is analytical and at the leading PM order behaves as $\sim 1+\mathcal{O}(\omega)$ for $L>0$, and $\sim \omega^{2|L|}$ for $L<0$.
Hence, in the $L$-sum, only the terms with $L>0$ contribute to poles of all order, while terms with $L<0,\, 2|L|<\ell$ contribute only to higher order poles, and terms with $2L<-\ell$ are analytical. 

We focus on the $\ell=2$ case, and compute the contribution of the poles in Eq.~\eqref{eq:tildeG1_prompt} to the time-domain GF up to $\mathcal{O}[\left(M/r'\right)^3]$. 
Since the equations are quite cumbersome, we show in the main text the explicit expressions only at first order $\mathcal{O}\left(M/r'\right)$, higher orders can be found in Appendix~\ref{app:p_coeffs}.
We expand Eq.~\eqref{eq:tildeG1_prompt} in the PM limit $\omega M,\, M/r'\ll1$, but keeping the prefactor $e^{-i\omega(t-r_*-t'+r')}$ to account for generic retarded times of the observer $\tau\equiv t-r_*$.
We anti-transform the result of this expansion, along the real $\omega$ axis. Since the integrand is singular in $\omega M=0$, the integral is computed through analytical continuation: by shifting the integration axis of a small quantity $\epsilon$ along the imaginary axis, which is positive due to causality.
To prevent the integrand from exponentially diverging at large $|\omega M|$, we close the contour on the lower-half plane for $t-r_*-t'+r'>0$, and on the upper-half plane for $t-r_*-t'+r'<0$. 
A schematization of the contour can be found in Fig.~\ref{fig:contour_plot_prompt_branch}.
Then, due to the exponential $e^{-i\omega(t-r_*-t'+r')}$, the contribution coming from the high-frequency arc is always zero following Jordan's lemma.
The only non-zero contribution is present when the contour is closed on the lower-half plane, i.e. for  $t-r_*-t'+r'>0$, and is given by the poles in $\omega M=0$.
The residue of each pole, at each order, is the coefficient of the of the $\omega^{-1}$ term in the Taylor expansion of the integrand in the $\omega\rightarrow 0$ limit.
The result it
\begin{equation}
\begin{split}
G^{(1),\,\rm Prompt}_{\ell=2 \,m,\,r_*\gg r_*'}&(t-t';r_*,r_*')=
\theta(u(t,r_*)-\bar{u}(t',r'))\times\\
&\left\lbrace\left[\frac{1}{2}  - \frac{3}{2r'}(u(t,r_*)-\bar{u}(t',r_*'))+ \frac{3}{4(r')^2} (u(t,r_*)-\bar{u}(t',r_*'))^2\right] -
\left[ \frac{11 - 6 \gamma_E - 12\log(2)}{2r'}+\right.\right.\\
&\left.\left.
\frac{-5 + 6 \gamma_E + 12\log(2)}{2 (r')^2} (u(t,r_*)-\bar{u}(t',r_*'))
-
\frac{5}{4 (r')^3}(u(t,r_*)-\bar{u}(t',r_*'))^2\right] \frac{M}{r'} + \mathcal{O}\left[\left(\frac{M}{r'}\right)^2\right]\right\rbrace\, ,
\end{split}
\end{equation}
where we recall that $u(t,r_*)\equiv t-r_*,\, \bar{u}(t',r')\equiv t'-r'$ and $\gamma_E$ is the Euler's constant.
At zeroth PM order, this prediction is consistent with the flat spacetime result in Eq.~\eqref{eq:G1_G2_results_flat}. Higher PM corrections do not modify the functional time-dependence of the prompt on the observer's retarded time. In fact, this can still be written as a polynomial of order $\ell=2$, containing all lower order terms up to the constant
\begin{equation}
\begin{split}
G^{(1),\,\rm Prompt}_{\ell=2\,m,\, r_*\gg r_*'}&(t-t';r_*,r_*')=\theta(u(t,r_*)-\bar{u}(t',r'))\times\\
&\left[p_0(r')+p_1(r')\, (u(t,r_*)-\bar{u}(t',r'))+p_2(r')\, (u(t,r_*)-\bar{u}(t',r'))^2\right]\, .
\end{split}
\label{eq:G1prompt_symbolic}
\end{equation}
The explicit expression of the coefficients $p_0,\,p_1,\,p_2$ with corrections up to $\mathcal{O}[(M/r')^3]$ is shown in Appendix~\ref{app:p_coeffs}.
We stress that we are working in a PM limit, hence we will obtain a result that is valid only for compact sources the limit of $r'/M\gg1$. Such sources excite small frequencies, rather than large ones, so that the error coming from generalizing a PM expansion to the whole complex plane can be neglected.
\subsection{Computation of $G_{\ell m}^{(2)}$}
\label{subsec:G2}
We now focus on computing $G^{(2)}_{\ell m,,r_*>r_*'}$, as defined in Eq.~\eqref{eq:G2_def}, in a PM expansion following the same procedure used for the computation of $G^{(1)}_{\ell m,,r_*>r_*'}$. 
As evident in Eq.~\eqref{eq:G2_def}, we first need to compute the ratio $A_{\rm out}/A_{\rm in}$. In analogy to the flat spacetime case, these coefficients can be found by expanding in the limit $r \rightarrow \infty$ the homogeneous solution $u^{\rm in}_\ell (\omega, r)$ defined in eq.~\eqref{eq:uIN_coulomb_superposition}, and matching it to  eq.~\eqref{eq:u_in_vs_u_out_u_inft-} expanded in the same limit,  using the asymptotic boundary conditions of $u^{\rm up,\infty-}$ as defined in Eqs.~\eqref{eq:u_out_asympt_expr} and~\eqref{eq:u_inf-_asympt_expr}.
This procedure yields
\begin{equation}
A_{\rm in}(\omega)=\mathcal{N}^{\rm in}e^{\pi\omega M} 2^{-4i\omega M-1}(-i\omega M)^{-\nu-2i\omega M}(\omega M)^{\nu}\sum_{L=-\infty}^{\infty}a_L\, i^{L+1}\frac{\left[\Gamma(\nu+L+1-2i\omega M)\Gamma(\nu+L+1+2i\omega M)\right]^{1/2}}{\Gamma(2L+\nu+1-2i\omega M)} \, ,
\label{eq:Ain}
\end{equation}
\begin{equation}
A_{\rm out}(\omega)=\mathcal{N}^{\rm in}e^{\pi\omega M} 2^{4i\omega M-1}(i\omega M)^{-\nu+2i\omega M}(\omega M)^{\nu}\sum_{L=-\infty}^{\infty}a_L\, (-i)^{L+1}\frac{\left[\Gamma(\nu+L+1-2i\omega M)\Gamma(\nu+L+1+2i\omega M)\right]^{1/2}}{\Gamma(2L+\nu+1+2i\omega M)} \, .
\label{eq:Aout}
\end{equation}
In first PM order and focusing on the multipole $\ell=2$, the ratio $A_{\rm out}/A_{\rm in}$ has both a single and a multi-valued contribution, and can be written as
\begin{equation}
\frac{A_{\rm out}(\omega)}{A_{\rm in}(\omega)}=-1-\frac{2i}{3} \left[-10 + 6 \gamma_E  + 12 \log 2 + 
   6\log(\omega M)\right] \omega M+\mathcal{O}[(\omega M)^2]\, .
   \label{eq:AoutAin_PM}
\end{equation}
We now compute $G^{(2)}_{\ell m,,r_*>r_*'}$ by inserting eqs.~\eqref{eq:Ain}, ~\eqref{eq:Aout}, ~\eqref{eq:u_up_coulomb_superposition} ~\eqref{eq:Hpm_vs_irregular_confluent_hypergeo} and~\eqref{eq:Kummer_transformation_uinfm_uup}, inside the definition in Eq.~\eqref{eq:G2_def}. This yields
\begin{equation}
\begin{split}
\tilde{G}^{(2)}&(\omega;r_*,r_*')=M\frac{A_{\rm out}}{A_{\rm in}}e^{-i\omega(t-r_*-t'-r')}A^{-2i\omega M}(r) \,\times \left\lbrace\right.
\\
& \left.\sum_{L=-\infty}^{\infty} a_L(\omega)\,(\omega M)^{L}\, \left[(-i\omega M)^{-1-2i\omega M+\nu}(\omega M)^{\nu+1}\right]\left(\frac{r'}{M}\right)^{L+\nu+1}\mathcal{A}^+_{L,\nu}(\omega)\, \mathrm{M}(L+\nu+1-2i\omega M,2L+2\nu+2,-2i\omega r') + \right.\\
&\left. \sum_{L=-\infty}^{\infty} a_L(\omega)\,(\omega M)^{-L} \, \left[(-i\omega M)^{-2-2i\omega M-2L-2\nu}(\omega M)^{1+2 L+\nu}\right]\left(\frac{r'}{M}\right)^{-L-\nu}\, \mathcal{B}^+_{L,\nu}(\omega)\, \mathrm{M}(2i\omega M-L-\nu,-2L-2\nu,-2i\omega r')\right\rbrace\, ,
\end{split}
\label{eq:tildeG2_vs_kummer_fun}
\end{equation}
where $\mathcal{A}^+, \mathcal{B}^+$ are defined as:
\begin{equation}\label{eq:AminusBminus}
\begin{split}
&\mathcal{A}^+_{L,\nu}(\omega)\equiv \tilde{N}^{\rm{up}}(\omega)\, C_{L,\nu}^+(\omega)\, 2^{L+\nu}\,  \frac{\pi}{\sin\left[2\pi(L+\nu+1)\right]}\,\frac{1}{\Gamma(2i\omega M-L-\nu)\Gamma(2L+2\nu+2)}\, , \\
&\mathcal{B}^+_{L,\nu}(\omega)\equiv \tilde{N}^{\rm{up}}(\omega)\, C_{L,\nu}^+(\omega)\, 2^{-L-\nu-1}\,  \frac{\pi}{\sin\left[2\pi(L+\nu+1)\right]}\,\frac{1}{\Gamma(L+\nu+1+2i\omega M)\Gamma(-2L-2\nu)}
\, .
\end{split}
\end{equation}
The PM structure of  $G^{(2)}_{\ell m,,r_*>r_*'}$ as written above is similar\footnote{The extra factor, $A_{\rm out}/A_{\rm in}$ does not alter the structure of the equation in the PM limit as $A_{\rm out}/A_{\rm in}\approx -1+\mathcal{O}(\omega M)$, see Eq.~\eqref{eq:AoutAin_PM}.} 
to that of $G^{(1)}_{\ell m,,r_*>r_*'}$ in Eq.~\eqref{eq:tildeG1_vs_kummer_fun}.
Working in this limit, it is trivial to show by similar arguments that the second line features a multi-valued contribution, generating a branch cut starting from $\omega=0$ along the negative imaginary axis, and a single-valued one analytical over the whole complex plane. Since we are not interested in tails, we will discard the contribution of the second line of Eq.~\eqref{eq:tildeG2_vs_kummer_fun}.
The PM expansion of the third line in Eq.~\eqref{eq:tildeG2_vs_kummer_fun}, instead, decouples a branch cut feature from a single-valued contribution with poles in $\omega=0$, of order up to $\mathcal{O}(\ell+1)$, exactly as discussed in Sec.~\ref{subsec:G1} for $\tilde{G}^{(1)}$, and as shown in the schematic representation of Fig.~\ref{fig:contour_plot_prompt_branch}.
When moving to the time domain, the residues at the poles give rise to a polynomial in $(u(t,r_*)-\bar{v}(t',r'))$, while the high-frequency arc contribution vanishes due to Jordan's lemma. 
In fact, the exponential $e^{-i\omega (t-t'+r_*+r')}$ requires the arc to be closed on the lower-half plane when $t-t'+r_*+r'>0$ and on the upper-half one otherwise, with the pole being swept only in the former case due to causality.
\begin{equation}
\begin{split}
G^{(2),\rm Prompt}_{\ell=2 \, m,\,r_*\gg r_*'}&(t-t';r_*,r_*')=
\theta(u(t,r_*)-\bar{v}(t',r'))\times\\
&\left\lbrace\left[-\frac{1}{2}  - \frac{3}{2r'}(u(t,r_*)-\bar{v}(t',r'))- \frac{3}{4(r')^2} (u(t,r_*)-\bar{v}(t',r'))^2\right] -
\left[
\frac{9 - 6 \gamma_E - 12 \log(2)}{2r'}+
 \right. \right. \\
&\left.\left.
\frac{15 - 6 \gamma_E - 12 \log(2)}{2(r')^2}(u(t,r_*)-\bar{v}(t',r'))+
\frac{5}{4 (r')^3}(u(t,r_*)-\bar{v}(t',r'))^2\right] \frac{M}{r'} + \mathcal{O}[\left(M/r'\right)^2]\right\rbrace
\end{split}
\end{equation}
At zeroth PM order, we have re-obtained the flat spacetime result in Eq.~\eqref{eq:G1_G2_results_flat}. 
At generic PM order, we see that also $G^{(2)}$ can be written as a polynomial of order $\ell=2$, containing all lower order terms up to the constant
\begin{equation}
\begin{split}
G^{(2),\rm Prompt}_{\ell=2 \, m,\,r_*\gg r_*'}&(t-t';r_*,r_*')=\theta(u(t,r_*)-\bar{v}(t',r'))\times\\
&\left[q_0(r')+q_1(r')\, (u(t,r_*)-\bar{v}(t',r'))+q_2(r')\, (u(t,r_*)-\bar{v}(t',r'))^2\right]\, .
\end{split}
\label{eq:G2prompt_symbolic}
\end{equation}
The explicit expression of the coefficients $q_0,\,q_1,\,q_2$ with corrections up to $\mathcal{O}[(M/r')^3]$ is shown in Appendix~\ref{app:q_coeffs}.
\subsection{Prompt response expression and cancellation among $G_{\ell m}^{(1)}$ and $G_{\ell m}^{(2)}$}

To obtain the full expression of the prompt response in the time domain, we sum the two contributions $G^{(1)},\,G^{(2)}$ computed in the previous sections, as shown in Eqs.~\eqref{eq:G1prompt_symbolic} and~\eqref{eq:G2prompt_symbolic}.
The contribution $G^{(1)}$ is activated after a time $t-r_*>t'-r'$, sooner than $G^{(2)}$, which contributes after $t-r_*>t'+r'$. Interestingly, we verified that after $t-r_*>t'+r'$ the two contributions cancel completely in the PM expansion up to $\mathcal{O}[(\omega M)^3]$, for $\ell=2$.
Hence, the PM expansion of the time-domain prompt response of the $\ell=2$ multipole, is computed as
\begin{equation}
\begin{split}
G^{\rm prompt}_{\ell=2\, m,\, r_*\gg r_*'}(t-t';r_*,r_*')&=\theta\left(u(t,r_*)-\bar{u}(t',r')\right)\theta\left(\bar{v}(t',r')-u(t,r_*)\right)\times\\
&\left[p_0(r')+p_1(r')\,\left(u(t,r_*)-\bar{u}(t',r')\right)+p_2(r')\, \left(u(t,r_*)-\bar{u}(t',r')\right)^2\right]\, .
\end{split}
\label{eq:Gprompt_PMexpansion}
\end{equation}

\section{Comparison with numerical solutions}
\label{sec:results}

\begin{figure*}[h!]
\includegraphics[width=0.99\textwidth]{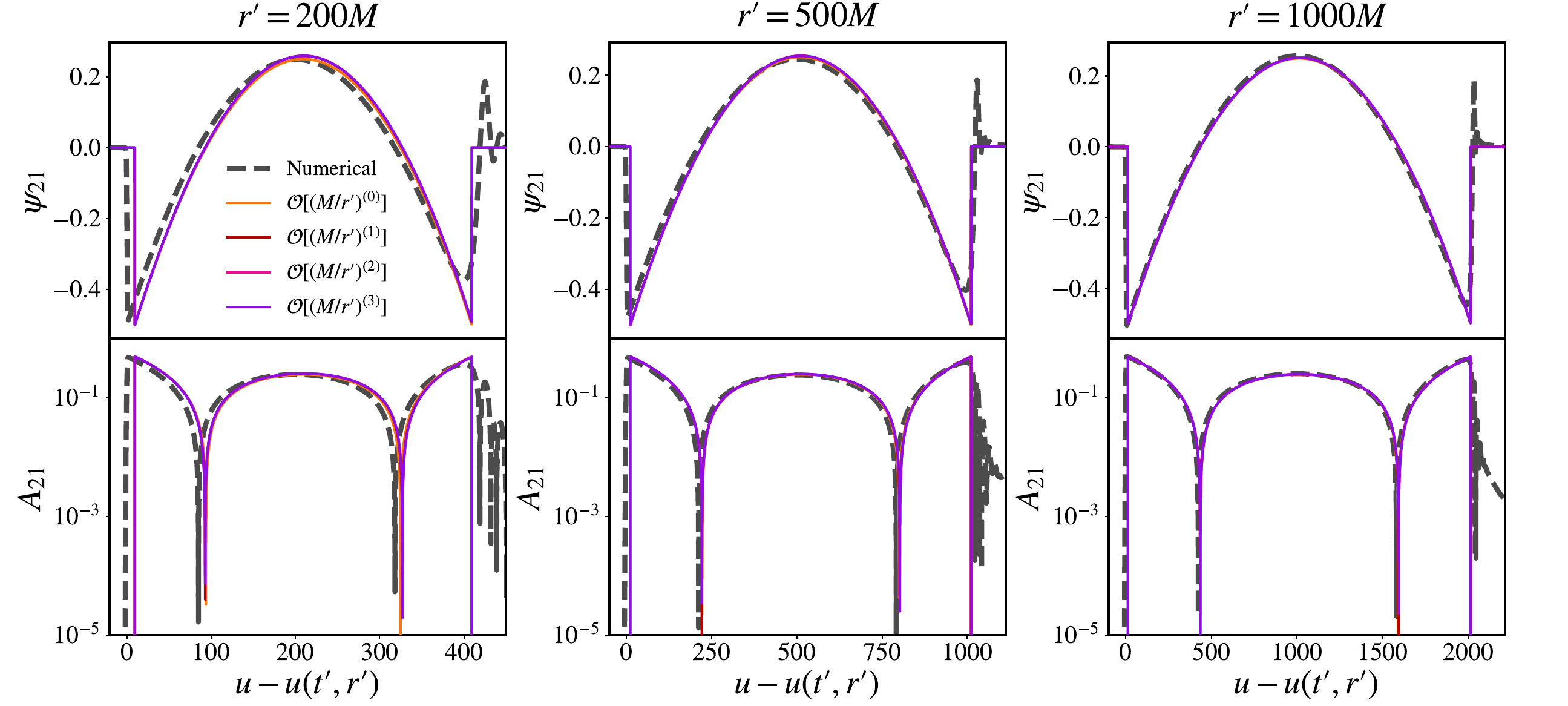}
\caption{Real component ('$+$' polarization) (\textbf{top}) and amplitude (\textbf{bottom}) of the Green's function $\ell=2,\,m=1$ multipole as observed at $\mathcal{I}^+$ vs the retarded time of the observer $u$. This quantity has been rescaled with the retarded time at which the first signal travelling on the curved light cone would arrive to $\mathcal{I}^+$.
Each column shows results for different location of the initial impulse $r'$, as displayed by the titles on the top axes.
In dashed black, results of numerical integrations obtained through the \textsc{RWZHyp} code~\cite{Bernuzzi:2010ty,Bernuzzi:2011aj}, as described in Sec.~\ref{sec:background}.
In color, the prediction obtained through a PM expansion up to $\mathcal{O}[(M/r')^n]$, with $n$ shown in the legend.
\label{fig:OG3_largeR}}
\end{figure*}
\begin{figure*}[h!]
\includegraphics[width=0.99\textwidth]{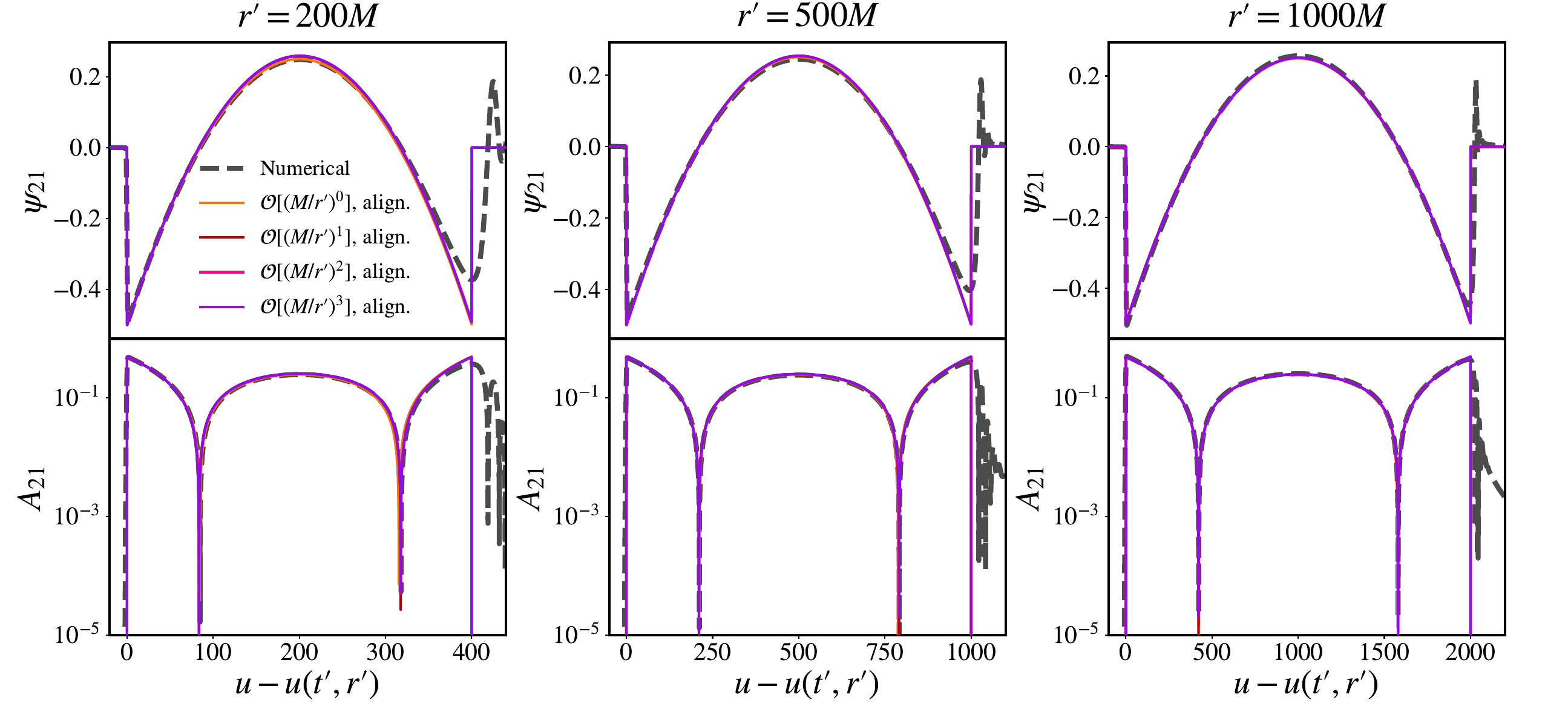}
\caption{Same as in Fig.~\ref{fig:OG3_largeR}, but the PM predictions have been shifted in the retarded time $u$ to minimize the  mismatch with the numerical waveform, as discussed in the text.
\label{fig:OG3_largeR_aligned}}
\end{figure*}
\begin{figure*}[h!]
\includegraphics[width=0.5\textwidth]{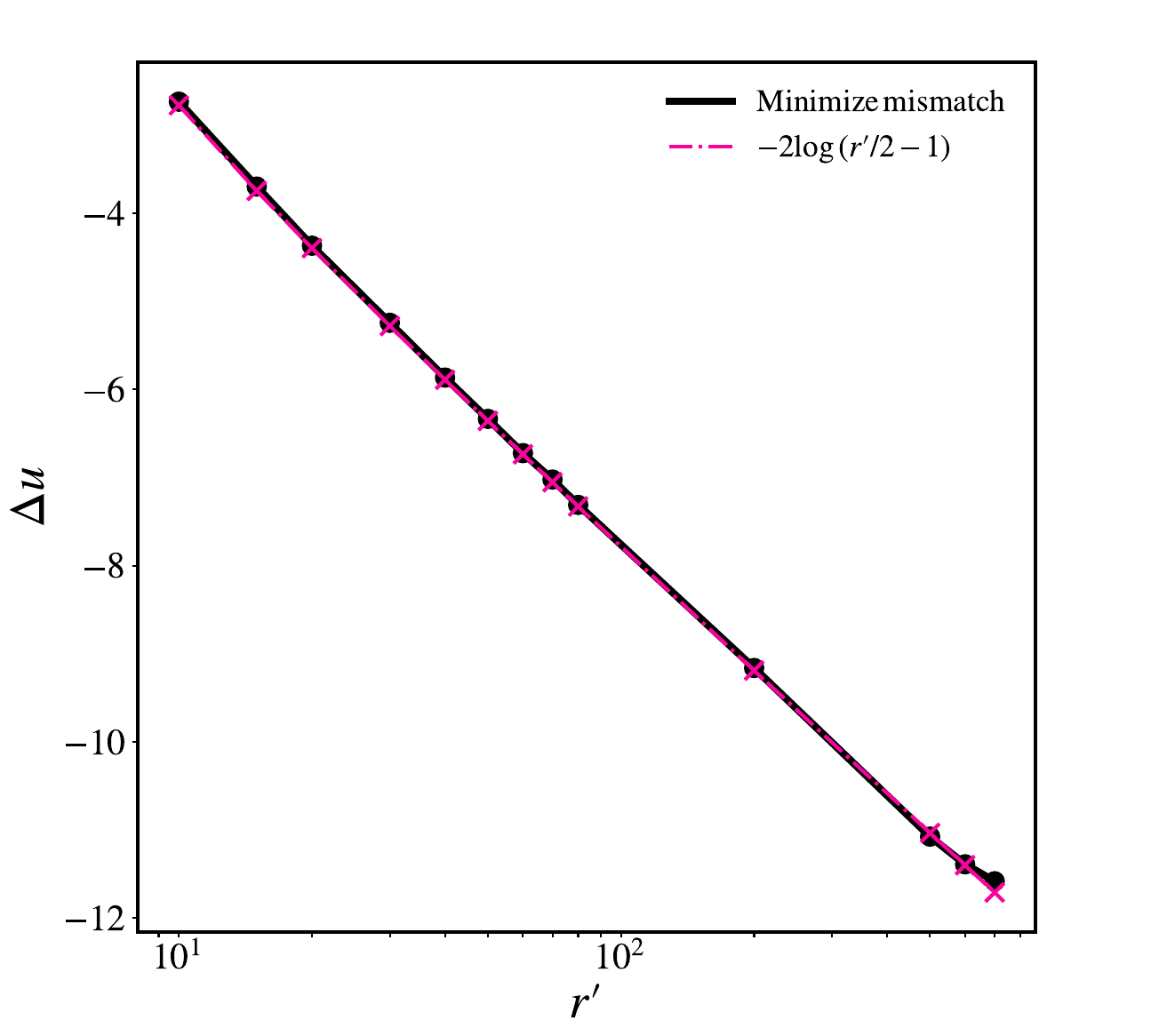}
\caption{In black, shift in the retarded time $u$ of the observer at $\mathcal{I}^+$ needed to minimize the mismatch between the numerical waveform and the predictions, i.e. to pass from results in Fig.~\ref{fig:OG3_largeR} to Fig.~\ref{fig:OG3_largeR_aligned}, vs $r'$.
\label{fig:delta_T_ell2em1}}
\end{figure*}

We test the analytical predictions for the prompt response obtained in the previous section, Eq.~\eqref{eq:Gprompt_PMexpansion}, against numerical computations of the time-domain GF.
For details regarding the numerical evolutions, we refer to Sec.~\ref{sec:background} and references therein.
We focus on the odd multipole $\ell=2,\,m=1$.
In Fig.~\ref{fig:OG3_largeR}, we show this comparison for different PM orders up to $\mathcal{O}[(M/r')^3]$ and three different locations of the initial impulse $r'=200M,\,500M,\,1000M$.
We observe that the PM corrections have a small impact on the predicted waveform and do not improve the matching with the numerical waveform.
Moving from large distances, e.g., $r'=1000M$, to intermediate ones, e.g., $r'=200M$, there is a progressive dephasing among the predictions and the numerical evolutions.
To investigate the reason behind this dephasing, we have aliged the waveforms, shifting the prediction's time axis such that the mismatch is minimized, in a time window close to the starting time of the prompt response\footnote{To align the waveforms by minimizing the mismatch, we exclude later times in the window during which the prompt response is activated, in order to avoid possible contamination with the ringdown.}.
As shown in Fig.~\ref{fig:OG3_largeR_aligned}, the retarded time shift $\Delta u$ greatly improves the agreement. We have investigated the trend of $\Delta u$ for different initial positions of the impulse $r'$, see Fig.~\ref{fig:delta_T_ell2em1}. This quantity is well approximated as $\Delta u\approx-2\log(r'/2-1)$: the dephasing in Fig.~\ref{fig:OG3_largeR} is a consequence of the predictions always traveling on the flat lightcone, while the ``true'' (numerical) prompt response travels on the curved lightcone.
The reason behind this problem follows from the Coulomb expansion itself, and from the PM expansion of the source. In fact, both the causality-induced Heaviside function and the retarded-time shift $\bar{u}(t',r')$ in Eq.~\eqref{eq:Gprompt_PMexpansion}, come from the exponentials in Eq.~\eqref{eq:tildeG1_prompt} and Eq.~\eqref{eq:tildeG2_vs_kummer_fun}. These exponentials, when the solution is written in terms of Coulomb wave functions expansions, are $\exp\left\lbrace-i\omega [t-r_*-(t'\mp r'+2M\log A(r'))]\right\rbrace$. Moreover, after the PM expansion only the exponential $\exp\left\lbrace-i\omega[t-r_*\mp (t'-r')]\right\rbrace$ remains in the integrand. 
Instead, propagation on the curved light cone requires $\exp\left\lbrace-i\omega[t-r_*-(t'\mp r'\mp 2M\log (r'/2-1))]\right\rbrace$ as an exponential in the integrand. 

\subsection{Phenomenological models}
\begin{figure*}[h!]
\includegraphics[width=0.99\textwidth]{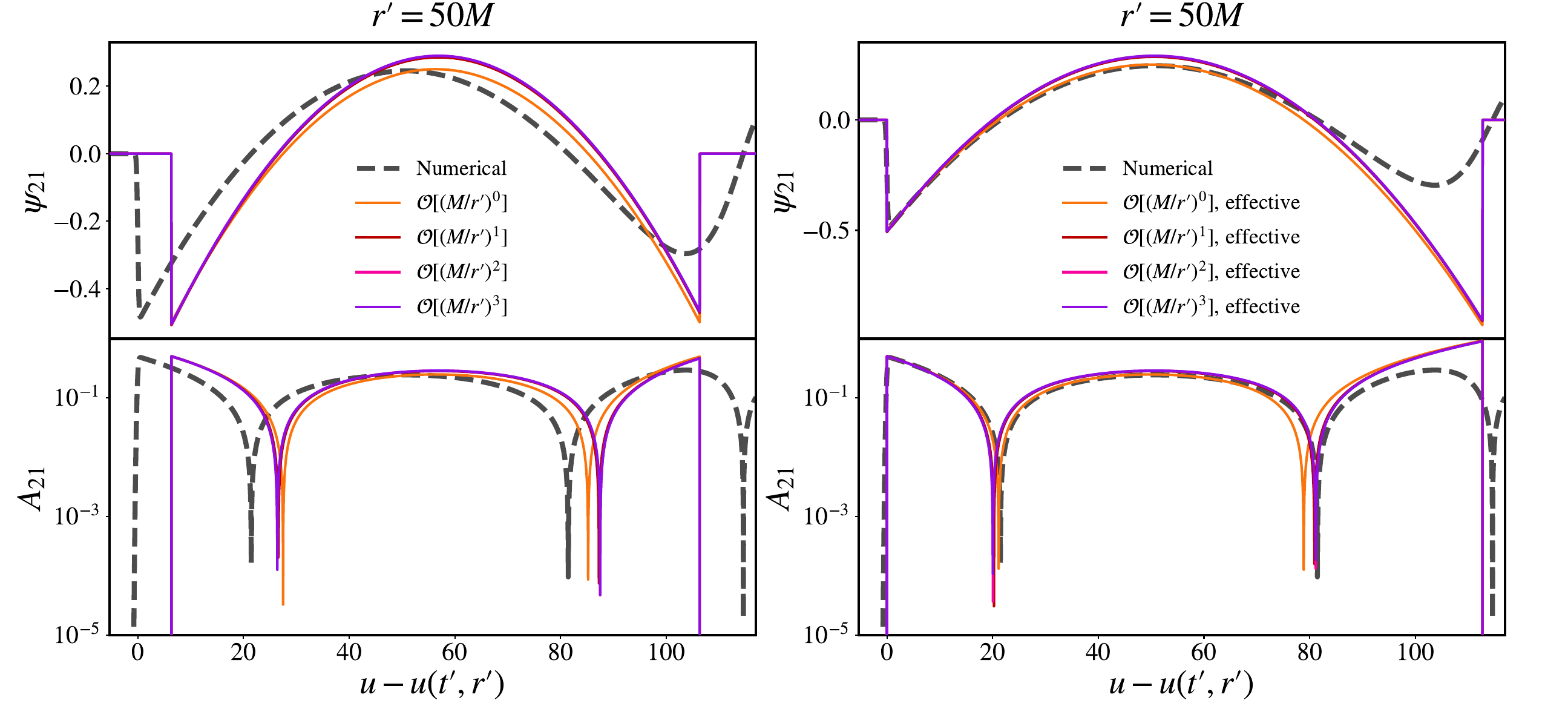}
\caption{Same as in Fig.~\ref{fig:OG3_largeR}, but the PM predictions have been shifted in the retarded time $u$ to minimize the  mismatch with the numerical waveform, as discussed in the text.
\label{fig:OG3_smallR_aligned_a_mano}}
\end{figure*}

Given the above discussion, we modify the prediction in Eq.~\eqref{eq:Gprompt_PMexpansion}, by shifting the (flat) retarded coordinates of the source $\bar{u}(t',r'),\,\bar{v}(t',r')$ to the curved retarded coordinates $\bar{u}(t',r')\rightarrow u(t',r_*'),\,\bar{v}(t',r')\rightarrow v(t',r_*')$. In doing so, we define a semi-phenomenological model of the prompt as
\begin{equation}
\begin{split}
G^{\rm prompt \, phenom.}_{\ell=2\, m,\, r_*\gg r_*'}(t-t';r_*,r_*')&=\theta\left(u(t,r_*)-u(t',r_*')\right)\theta\left(v(t',r_*')-u(t,r_*)\right)\times\\
&\left[p_0(r')+p_1(r')\,\left(u(t,r_*)-u(t',r_*')\right)+p_2(r')\, \left(u(t,r_*)-u(t',r_*')\right)^2\right]\, .
\end{split}
\label{eq:promptG_phenom} 
\end{equation}
In Fig.~\ref{fig:OG3_smallR_aligned_a_mano} we test this prediction for intermediate distances, in particular $r'=50M$, and compare it with the first-principles result in Eq.~\eqref{eq:Gprompt_PMexpansion}. 
As expected, the time shift added in Eq.~\eqref{eq:promptG_phenom} eliminates the dephasing observed in Fig.~\ref{fig:OG3_largeR}, already at the $\mathcal{O}[(M/r')^0]$ level. 
However, while the PM series appears to converge already at $\mathcal{O}[(M/r')^3]$, PM corrections do not improve the match between the semi-phenomenological model in Eq.~\eqref{eq:promptG_phenom} and the numerical prompt response.
This hints that the solution written in terms of Coulomb wave functions is only suited for very large distances.
It is indeed well known in the literature that this expansion does not converge at the horizon~\cite{Mano:1996vt,Mano:1996mf,Sasaki:2003xr}, however, it is not clear at which distances the convergence becomes poor.
\\

\begin{figure*}[h!]
\includegraphics[width=0.99\textwidth]{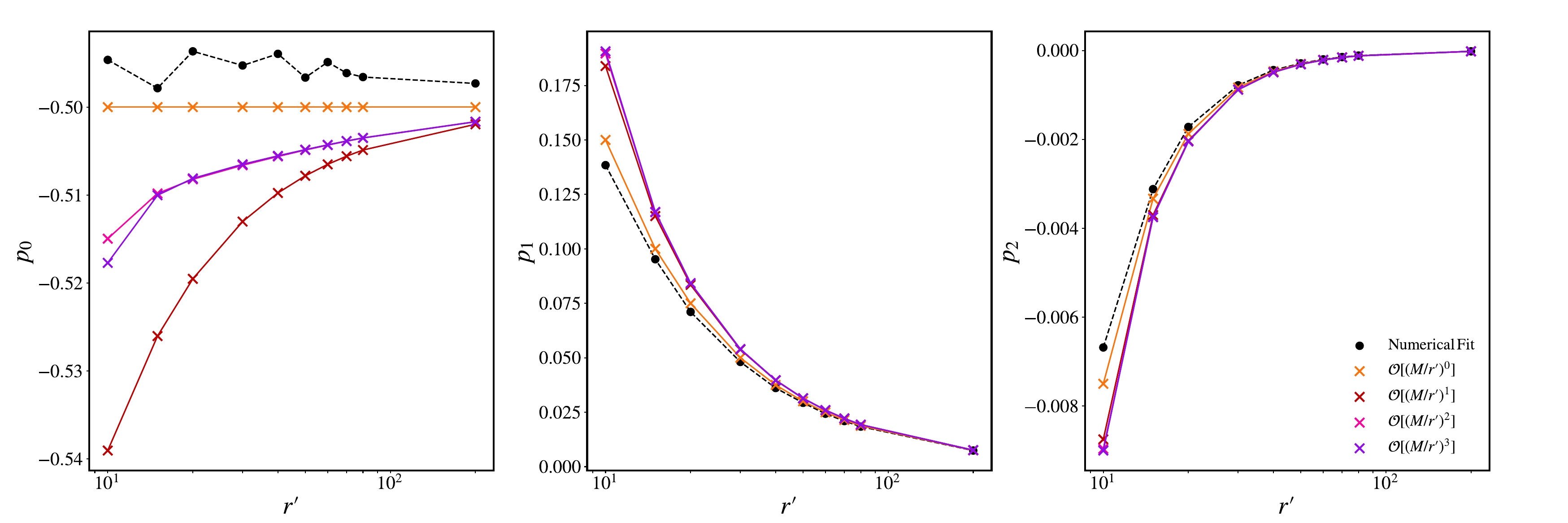}
\caption{Coefficients of the model Eq.~\eqref{eq:promptG_fit} obtained fitting against numerical evolutions (black dot), and analytical values of the coefficients in the first-principles prediction Eq.~\eqref{eq:Gprompt_PMexpansion} at different PM order (crosses) vs the location $r'$ of the impulsive source.
\label{fig:coeff_num_vs_fit_ell2em1}}
\end{figure*}

\begin{figure*}[h!]
\includegraphics[width=0.99\textwidth]{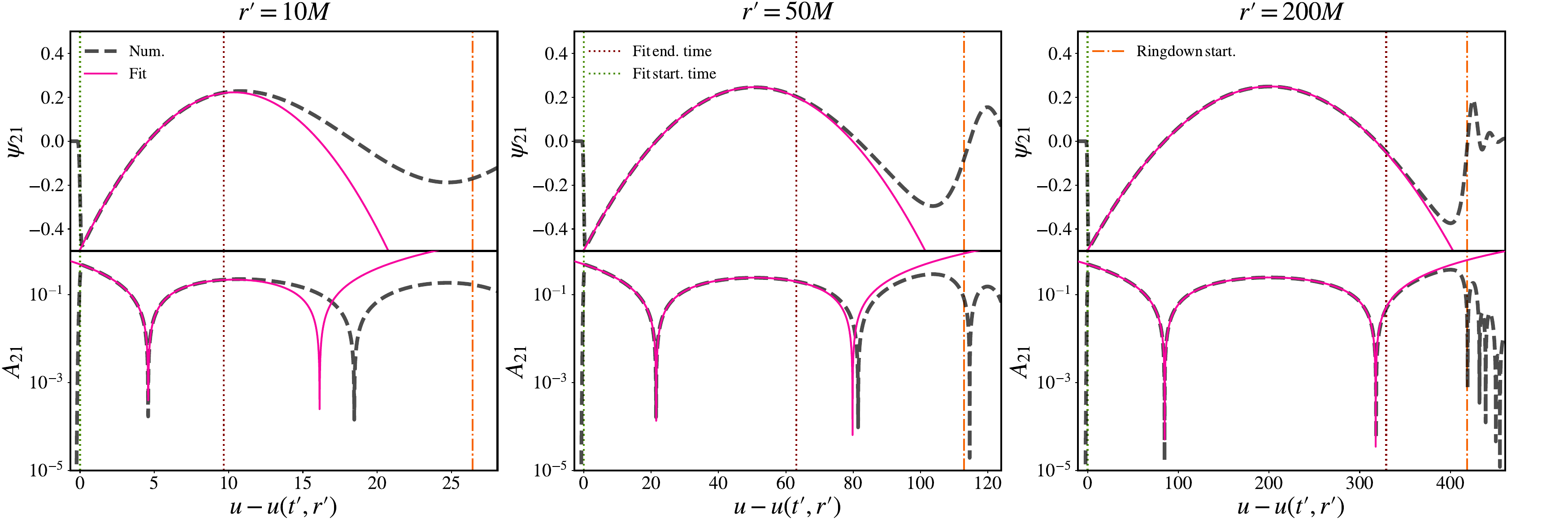}
\caption{Real component ('$+$' polarization) (\textbf{top}) and amplitude (\textbf{bottom}) of the Green's function $\ell=2,\,m=1$ multipole as observed at $\mathcal{I}^+$ vs the retarded time of the observer $u$. This quantity has been rescaled with the retarded time at which the first signal travelling on the curved light cone would arrive to $\mathcal{I}^+$.
Each column shows results for different location of the initial impulse $r'$, as displayed by the titles on the top axes.
In dashed black, results of numerical integrations obtained through the \textsc{RWZHyp} code~\cite{Bernuzzi:2010ty,Bernuzzi:2011aj}, as described in Sec.~\ref{sec:background}.
In purple, the prompt-response waveform obtained fitting with the model Eq.~\eqref{eq:promptG_fit}. 
The parameters of the fit, for different $r'$, are shown in Fig.~\ref{fig:coeff_num_vs_fit_ell2em1}.
Horizontal lines represent: starting time of the fit (green dotted), ending time of the fit (red dotted), and ringdown starting time (orange dot-dashed).
\label{fig:wf_abs_num_vs_fit_ell2em1_different_r0}}
\end{figure*}

To further investigate the regime of intermediate and small distances $r'$, we have fitted the numerical prompt response with a polynomial expression of order $n=2$ in $(u(t,r_*)-u(t',r_*'))$. Then, we have compared the fitted values of $p_i$ with the coefficients in Eq.~\eqref{eq:promptG_phenom}.
The result of this experiment is shown in Fig.~\ref{fig:coeff_num_vs_fit_ell2em1}, for various $r'$.
Products of the fit, tested against the numerical prompt response can be found in Fig.~\ref{fig:wf_abs_num_vs_fit_ell2em1_different_r0} for $r'=10M,\,50M,\,200M$.
As the initial position of the pulse $r'$ is closer to the black hole, the time window during which the prompt response can be approximated as a polynomial is progressively shorter, so that this model holds only at the start of the prompt response and fails at later times, close to the ringdown. 
As a consequence, also the time window used to fit the coefficients $p_i$ has been adjusted with $r'$, as shown in Fig.~\ref{fig:wf_abs_num_vs_fit_ell2em1_different_r0}, so that the model we fit is
\begin{equation}
\begin{split}
G^{\rm prompt \, fit}_{\ell=2\, m,\, r_*\gg r_*'}(t-t';r_*,r_*')&=\theta\left(u(t,r_*)-u(t',r_*')\right)\theta\left(t_{\rm end}(r_*')-u(t,r_*)\right)\times\\
&\left[p_0^{\rm fit}(r')+p_1^{\rm fit}(r')\,\left(u(t,r_*)-u(t',r_*')\right)+p_2^{\rm fit}(r')\, \left(u(t,r_*)-u(t',r_*')\right)^2\right]\, .
\end{split}
\label{eq:promptG_fit}
\end{equation}
with $p_i^{\rm fit}$ as free parameters and $t_{\rm end}$ as phenomenological parameter to fix before the fitting procedure starts.
Even though Eq.~\eqref{eq:promptG_fit} can fit a portion of the signal with great accuracy, as shown in Fig.~\ref{fig:wf_abs_num_vs_fit_ell2em1_different_r0},
this is not sufficient to claim that a polynomial of order $n=2$ is present in the prompt response for small and intermediate distance: the model could be overfitting the data.
However a strong indication that Eq.~\eqref{eq:promptG_fit} reflects the ``real'' structure of the prompt response, to some extent, comes from the comparison in Fig.~\ref{fig:coeff_num_vs_fit_ell2em1}.
For large $r'$, the fitted coefficients $p_i^{\rm fit}$ are in good agreement with those of the semi-phenomenological model in Eq.~\eqref{eq:promptG_phenom}, except for a constant shift that we address as potential numerical error.
As $r'$ decreases, the coefficients $p_{1,2}$ follow a similar trend and remain relatively close, for the $\mathcal{O}[(M/r')^0]$ case.
For higher PM corrections, instead, the fitted coefficient $p_0$ departs from the predicted one, as the source is located closer to the BH.
We argue that these results hint at the presence of a Eq.~\eqref{eq:promptG_fit}-like behavior in the prompt response. 
Moreover, they further confirm that the Coulomb-wave-function expansion used to compute the RWZ solutions, Eq.~\eqref{eq:psi_as_coulomb_superposition}, is not convergent at intermediate and small values of $r'$.
We elaborate more on these points in Sec.~\ref{sec:conclusions}
\\ \\ 

%

\section{Conclusions and future directions}   %
\label{sec:conclusions} %
%

%
In this work, we investigated the Green's function of a Schwarzschild black hole governing its linear response to an impulsive -- Dirac delta -- perturbation, in the case of an observer at $\mathcal{I}^+$.
We focused on the initial component of this response, the one traveling on the curved light cone and denoted as prompt response~\cite{Leaver:1986gd,Andersson:1996cm} in the past literature.
Historically, the prompt response was though to originate from the high-frequency arcs component of the Green's function in the complex-frequency, Laplace-transform domain.
We challenged this intuition by presenting an original analytical model of the prompt response, obtained by solving the Regge-Wheeler equation in the PM limit of large distances of the compact source, exciting the small-frequency response of the Schwarzschild background. 
By adopting a solution written as superposition of Coulomb wave functions, we showed that the prompt response originates from poles at $\omega =0$ in the Laplace-transform domain, of order $1\leq n\leq\ell+1$, with $\ell$ multipole number.
Once we integrate over the $\omega$-frequency plane and move back to the time domain, the residues at the poles give rise to a polynomial of order $\ell$ in the retarded time of the observer. 
We validated these predictions, see Eq.~\eqref{eq:Gprompt_PMexpansion}, against numerical evolutions of a narrow Gaussian source, located at different distances from the black hole, focusing on the $\ell=2,\,m=1$ multipole.
The predictions agree with numerical evolutions at very large distances, in the flat space-time limit.
A growing mismatch is present as the source is localized closer to the BH, characterized by a growing dephasing among predicted and numerical waveform, which is not resolved or improved by adding PM corrections.
We found that this dephasing is due to the prediction traveling on the flat light-cone, while the ``true'' prompt response travels on the curved one. We argue that this is a consequence of: the Coulomb wave function expansion we used as solution, which fails when the source is not at large distances; the PM expansion treating perturbatively corrections to the flat light-cone, while some sort or ``resummation'' appears necessary.
To compensate for this latter issue of our approach, we introduced a semi-phenomenological model of the prompt, by substituting the flat-spacetime retarded coordinates of the source with curved-spacetime ones; see Eq.~\eqref{eq:promptG_phenom}.
This model improves the mismatch with the numerical prompt response at large-intermediate distances; however, the improvement is marginal and it is restricted to the elimination of the dephasing.
To gain insight on how to further improve the phenomenological model, we fitted the numerical prompt response with a polynomial of order $n=2$. When compared with our semi-phenomenological model, this experiment hints at the presence of a polynomial-like component in the prompt, even for sources located at small distances $r'=10M$. However, the PM approach based on a Coulomb wave function expansion (Eq.~\eqref{eq:psi_as_coulomb_superposition}) is not converging to the real solution, and a different representation of the solution is needed.  
\\ 

%
It is known that the expansion in Coulomb wave functions in Eq.~\eqref{eq:psi_as_coulomb_superposition}, fails at the BH horizon. 
In the literature, solutions valid over the full parameter range $2M<r'<\infty$, well converging at small frequencies, are built by using this type of representation only at large distances, and matching it with one written as superposition of hypergeometric functions at small distances. This approach is called ``Mano-Suzuki-Takasugi'' (MST)~\cite{Mano:1996vt,Mano:1996mf,Sasaki:2003xr}.
We aim to investigate the prompt response with a solution written in terms of hypergeometric functions, working still in a small-frequencies limit $\omega \approx 0$, but now at small distances of the compact source as well.
Finally, in the present work we only restricted to the pole in zero, neglecting the contribution of the branch cut. In Refs.~\cite{Arnaudo:2025uos,Kuntz:2025gdq,Arnaudo:2025kit}, it was recently shown in the Schwarzschild-de Sitter (SdS) and P$\mathrm{\ddot{o}}$schl-Teller geometries, that the prompt response is determined by a series of poles along the imaginary axis. In the asymptotically flat limit, Ref.~\cite{Arnaudo:2025kit} argue that the SdS poles accumulate to form a branch cut originating Price's law. Further investigations are needed to understand if other components of the imaginary axis could affect the prompt. 
The computation done in this work focus on the Regge-Wheeler sector, which describes even multipoles. The extension to this sector is natural once the odd is solved, in the Schwarzschild case, and simply involves more cumbersome expressions that we leave for future work.
Since astrophysical black holes have spin, a natural and needed extension of our work is to investigate the prompt response of a Kerr black hole. The final aim is, in fact, to leverage the analytical understanding of the prompt in waveform modeling.
Current waveform models for binary black hole merger, lack a first-principle, closed-form expression for the near-peak transient leading to the stationary ringdown. 
An analytical expression for the prompt response could supplement the dynamical ringdown excitation picture of Ref.~\cite{DeAmicis:2025xuh}, to provide such a model in the extreme mass-ratio limit, where non-linearities are suppressed.
This could give phenomenological insight to the observationally-relevant comparable masses case, given the un-expected linear behavior of numerical simulations of such systems; see e.g. Refs.~\cite{DeAmicis:2024eoy,DeAmicis:2025xuh} and discussions therein.

\acknowledgments
M.D.A. is grateful to Marc Casals and Eric Poisson for useful discussions and pointers to relevant literature.
E.C. is thankful to Adrien Kuntz for useful conversations and comments on the manuscript.
We thank Gregorio Carullo and Laura Sberna for instructive discussions, relevant comments on the draft and careful guidance throughout the years.
E.C. acknowledges financial support
provided under the European Union’s H2020 ERC
Advanced Grant “Black holes: gravitational engines of
discovery” grant agreement no. Gravitas–101052587.
Views and opinions expressed are however those of the
author only and do not necessarily reflect those of the
European Union or the European Research Council.
Neither the European Union nor the granting authority
can be held responsible for them. E.C. also thank the
Fundação para a Ciência e Tecnologia (FCT), Portugal,
for the financial support to the Center for Astro-
physics and Gravitation (CENTRA/IST/ULisboa)
through grant No. UID/PRR/00099/2025
(https://doi.org/10.54499/UID/PRR/00099/2025)
and grant No. UID/00099/2025
(https://doi.org/10.54499/UID/00099/2025). 
E.C. also acknowledge
financial support by the VILLUM Foundation (grant
no. VIL37766) and the DNRF Chair program (grant
no. DNRF162) by the Danish National Research
Foundation. 
This research was supported in part by
Perimeter Institute for Theoretical Physics. Research at Perimeter Institute is supported in part by the Government of Canada through the Department of Innovation, Science and Economic Development and by the Province of Ontario through the Ministry of Colleges and Universities.
\\

%
\appendix
\section{Coulomb wavefunctions}
\label{app:Coulomb}
Consider the differential equation
\begin{equation}
\frac{d^2 u}{dz^2} + \left[ 1 - \frac{2\eta}{z} - \frac{\lambda(\lambda+1)}{z^2} \right] u = 0 \,,
\end{equation}
where \(z\in \mathbb{C}\) is the independent variable, \(\lambda\) is a fixed index, 
and \(\eta \in \mathbb{C}\) is a general complex parameter. This equation is a 
generalized Coulomb-type equation, and it admits two linearly independent solutions 
traditionally denoted by \(F_\lambda(\eta,z)\) and \(G_\lambda(\eta,z)\).

The function \(F_\lambda(\eta,z)\), referred to as the \emph{regular Coulomb function}, 
is characterized by its behavior near the origin, \(F_\lambda(\eta,z) \sim z^{\lambda+1}\) 
as \(z \to 0\). It can be expressed in terms of the Tricomi confluent hypergeometric function 
\(U\) as
\begin{equation}
F_\lambda(\eta,z) = C_\lambda(\eta)\, z^{\lambda+1} e^{-i z} \, 
U(\lambda+1 - i\eta, 2\lambda+2, 2 i z)\,,
\end{equation}
where \(C_\lambda(\eta)\) is a normalization constant depending on \(\lambda\) and \(\eta\).

The second independent solution, \(G_\lambda(\eta,z)\), called the \emph{irregular Coulomb function}, 
is linearly independent from \(F_\lambda\) and can similarly be expressed in terms of hypergeometric functions. 
Its leading behavior near the origin is singular for \(\lambda \neq -1/2\), \(G_\lambda(\eta,z) \sim z^{-\lambda}\) 
as \(z \to 0\).

From these two functions, one can define the \emph{outgoing} and \emph{ingoing} Coulomb wave 
functions \(H^\pm_\lambda(\eta,z)\) as linear combinations:
\begin{equation}
H^+_\lambda(\eta,z) = F_\lambda(\eta,z) + i G_\lambda(\eta,z), \qquad
H^-_\lambda(\eta,z) = F_\lambda(\eta,z) - i G_\lambda(\eta,z) \,.
\end{equation}
These functions are particularly convenient because they exhibit simple asymptotic behavior 
for large \(|z|\):
\begin{equation}
H^\pm_\lambda(\eta,z) \sim e^{\pm i z} \quad \text{as } |z| \to \infty \,.
\end{equation}

In terms of hypergeometric functions, the outgoing and ingoing waves can be expressed as
\begin{align}
H^+_\lambda(\eta,z) &= z^{\lambda+1} e^{-i z} \, U(\lambda+1-i\eta,2\lambda+2,2 i z),\\
H^-_\lambda(\eta,z) &= z^{\lambda+1} e^{i z} \, {}U(\lambda+1+i\eta,2\lambda+2,-2 i z) \,.
\end{align}

Finally, in the limit \(\eta \to 0\), the Coulomb wave functions reduce to the standard 
free-wave solutions. Specifically, one has
\begin{equation}
F_\lambda(0,z) = z\, j_\lambda(z), \qquad
G_\lambda(0,z) = - z\, y_\lambda(z), \qquad
H^\pm_\lambda(0,z) = z\, h^\pm_\lambda(z) \,,
\end{equation}
where \(j_\lambda(z)\) and \(y_\lambda(z)\) are the regular and irregular spherical Bessel 
functions, and \(h^\pm_\lambda(z) = j_\lambda(z) \pm i y_\lambda(z)\) are the spherical Hankel functions. 
%

\section{Explicit post-Minkowskian expressions or $\nu$ and $a_L$}
\label{app:aL_coeffs}

In this appendix, we provide the explicit expressions of the coefficients $\nu,\, a_{L=0,\pm1,\pm2,+3}$ in Eq.~\eqref{eq:psi_as_coulomb_superposition}, which were used to compute 
the results of this work.

\begin{align}
&\nu =\ell - \frac{2 (24 + 13\ell (1 + \ell) + 15 \ell^2 (1 + \ell)^2)}{
 \ell (1 + \ell) (-1 + 2 \ell) (1 + 2 \ell) (3 + 2 \ell)}(\omega M)^2+\mathcal{O}[(\omega M)^4] \, ,\\
&a_0 = 1 \, , \\
&a_1 = 
\frac{\left(\ell^2+2 \ell-3\right)}{(\ell+1)(2\ell+1)}\,\omega M
-\frac{4 i}{2 \ell+1}\,(\omega M)^2 + \Bigg[
\frac{\ell \left(-4 \ell^3-12 \ell^2+\ell+38\right)+25}
{2 (\ell+1)^2 (2 \ell+1)^2}\,\nu  \nonumber\\
&\qquad
+\frac{\ell^8-19 \ell^7-291 \ell^6-1453 \ell^5-3558 \ell^4
-4318 \ell^3-1576 \ell^2+1554 \ell+1260}
{(\ell+1)^3 (\ell+2) (2 \ell+1) (2 \ell+3)^2 (2 \ell+5)}
\Bigg](\omega M)^3 - \nonumber\\
&\quad
\frac{4 i \left(8 \ell^{10}+76 \ell^9+602 \ell^8+3201 \ell^7
+10697 \ell^6+23723 \ell^5+35852 \ell^4+37476 \ell^3
+27625 \ell^2+13890 \ell+3600\right)}
{\ell(\ell+1)^2 (\ell+2) (2\ell-1) (2 \ell+1)^3 (2 \ell+3)^2 (2 \ell+5)}
\,(\omega M)^4+\mathcal{O}[(\omega M)^5] \, , \\
&a_{-1}=\frac{\left(4-\ell^2\right)  }{2 \ell^2+l} \omega M-\frac{4 i}{2\ell+1}(\omega M)^2+\mathcal{O}[(\omega M)^3] \, ,\\
&a_2=\frac{\ell (\ell+4) \left(\ell^2+2 \ell-3\right)}{(\ell+1) (2 \ell+1) (2 \ell+3)^2}(\omega M)^2-\frac{4 i \left(\ell^2+3 \ell-2\right)}{4 \ell^3+12 \ell^2+11 \ell+3}(\omega M)^3+\mathcal{O}[(\omega M)^4]\, ,\\
&a_{-2}=\frac{\left(\ell^2-4\right) \left(\ell^2-2 \ell-3\right)}{(1-2 \ell)^2 \ell (2 \ell+1)}(\omega M)^2+\mathcal{O}[(\omega M)^3]\, ,\\
&a_3=\frac{\ell (\ell+4) (\ell+5) \left(\ell^2+2 \ell-3\right) }{3 (2 \ell+1) (2 \ell+3)^2 \left(2 \ell^2+9 \ell+10\right)}(\omega M)^3-\frac{4 i \left(3 \ell^4+24 \ell^3+44 \ell^2-16 \ell-15\right) }{3 (2 \ell+3)^2 \left(4 \ell^3+16 \ell^2+17 \ell+5\right)}(\omega M)^4+\mathcal{O}[(\omega M)^5]\, .
\label{eq:aLi_expr}
\end{align}

\section{$G^{(1)}_{\ell=2 \, m}$ post-Minkowskian expansion}
\label{app:p_coeffs}
In this appendix, we provide the explicit expressions of the polynomial coefficients $p_{0,1,2}$ appearing in Eq.~\eqref{eq:G1prompt_symbolic} at third post-Minkowskian order\footnote{3PM order means $\mathcal{O}[(r'/M)^3]$ for $p_0$, $\mathcal{O}[(r'/M)^4]$ for $p_1$ and $\mathcal{O}[(r'/M)^5]$ for $p_2$, since these terms are multiplied by contributions of order $\mathcal{O}[(r'/M)^0],\,\mathcal{O}[(r'/M)],\,\mathcal{O}[(r'/M)^2]$, respectively, in Eq.~\eqref{eq:G1prompt_symbolic}}.
\begin{align}
& p_0(r')= -\frac{i}{2}-\frac{i}{2}\left[-11 + 6 \gamma_E + 12\log(2)\right]\frac{M}{r'}-
\frac{i}{70}\left[
819 - 564 \gamma_E + 210 \gamma_E^2 - 35 \pi^2 + 840 \gamma_E \log(2) - 457 \log(4) +
\right.\nonumber\\
&\left.\qquad
 420 \log(2) \log(4) - 214 \log(r'/M)\right] \left(\frac{M}{r'}\right)^2 -
\frac{i}{630}\left[16252 - 8310 \gamma_E + 3150 \gamma_E^2 - 525 \pi^2 - 
  13410 \log(2) + \right.\nonumber\\
&\qquad\left.
12600 \gamma_E \log(2) + 12600 \log(2)^2 - 
  3210 \log(r'/M)\right]\left(\frac{M}{r'}\right)^3+\mathcal{O}\left[\left(\frac{M}{r'}\right)^4\right]\\  
&p_1(r') =\frac{3 i M }{2 r'}+\frac{i}{2} \left(-5 + 6 \gamma_E + \log(4096)\right)\left(\frac{M}{r'}\right)^2+ 
\frac{5i}{21}\left(-17 + 21 \gamma_E + 42 \log(2)\right)
\left(\frac{M}{r'}\right)^3 +\nonumber\\
&\qquad
\frac{5i}{14}\left(-19 + 24 \gamma_E + 48 \log(2)\right)
\left(\frac{M}{r'}\right)^4 +
\mathcal{O}\left[\left(\frac{M}{r'}\right)^5\right]
\\
&p_2(r')=-\frac{3 i}{4}\left(\frac{M}{r'}\right)^2 - \frac{5 i}{4}\left(\frac{M}{r'}\right)^3 - \frac{15 i }{7}\left(\frac{M}{r'} \right)^4 -
\frac{15 i}{4}\left(\frac{M}{r'}\right)^5 + \mathcal{O}\left[\left(\frac{M}{r'}\right)^6\right]\, .
\end{align}

\section{$G^{(2)}_{\ell=2 \, m}$ post-Minkowskian expansion}
\label{app:q_coeffs}
In this appendix, we provide the explicit expressions of the polynomial coefficients $q_{0,1,2}$ appearing in Eq.~\eqref{eq:G2prompt_symbolic} at third post-Minkowskian order\footnote{3PM order means $\mathcal{O}[(r'/M)^3]$ for $q_0$, $\mathcal{O}[(r'/M)^4]$ for $q_1$ and $\mathcal{O}[(r'/M)^5]$ for $q_2$, since these terms are multiplied by contributions of order $\mathcal{O}[(r'/M)^0],\,\mathcal{O}[(r'/M)],\,\mathcal{O}[(r'/M)^2]$, respectively, in Eq.~\eqref{eq:G2prompt_symbolic}}.
\begin{align}
& q_0(r')= \frac{i}{2}-\frac{i}{2}\left[-9 + 6 \gamma_E + 12\log(2)\right]\frac{M}{r'}+
\frac{i}{210}\left[
5957 - 3792 \gamma_E + 630 \gamma_E^2 - 105 \pi^2 - 
 6942 \log(2) + 2520 \gamma_E \log(2) + 
\right.\nonumber\\
&\left.\qquad
2520 \log(2)^2 - 642 \log(r'/M)
\right] \left(\frac{M}{r'}\right)^2 +
\frac{i}{630}\left[
34252 - 19110 \gamma_E + 3150 \gamma_E^2 - 525 \pi^2 - 
 35010 \log(2) + 
\right.\nonumber\\
&\qquad\left.
12600 \gamma_E \log(2) + 12600 \log(2)^2 - 
 3210 \log(r'/M)
\right]\left(\frac{M}{r'}\right)^3+\mathcal{O}\left[\left(\frac{M}{r'}\right)^4\right]\\  
&q_1(r') =\frac{3 i M }{2 r'}-\frac{i}{2} \left(-15 + 6 \gamma_E + 12\log(2)\right)\left(\frac{M}{r'}\right)^2- 
\frac{5i}{21}\left(-53 + 21 \gamma_E + 42 \log(2)\right)
\left(\frac{M}{r'}\right)^3 -\nonumber\\
&\qquad
\frac{5i}{14}\left(-61 + 24 \gamma_E + 48 \log(2)\right)
\left(\frac{M}{r'}\right)^4 +
\mathcal{O}\left[\left(\frac{M}{r'}\right)^5\right]
\\
&q_2(r')=\frac{3 i}{4}\left(\frac{M}{r'}\right)^2 + \frac{5 i}{4}\left(\frac{M}{r'}\right)^3 + \frac{15 i }{7}\left(\frac{M}{r'} \right)^4 +
\frac{15 i}{4}\left(\frac{M}{r'}\right)^5 + \mathcal{O}\left[\left(\frac{M}{r'}\right)^6\right]\, .
\end{align}

\bibliography{bibliography}

@article{Cardoso:2019rvt,
    author = "Cardoso, Vitor and Pani, Paolo",
    title = "{Testing the nature of dark compact objects: a status report}",
    eprint = "1904.05363",
    archivePrefix = "arXiv",
    primaryClass = "gr-qc",
    doi = "10.1007/s41114-019-0020-4",
    journal = "Living Rev. Rel.",
    volume = "22",
    number = "1",
    pages = "4",
    year = "2019"
}

@misc{NIST:DLMF,
         key = "{\relax DLMF}",
       title = "{\it NIST Digital Library of Mathematical Functions}",
howpublished = "\url{https://dlmf.nist.gov/}, Release 1.2.4 of 2025-03-15",
         url = "https://dlmf.nist.gov/",
        note = "F.W.J. Olver, A.B. {Olde Daalhuis}, D.W. Lozier, B.I. Schneider,
                R.F. Boisvert, C.W. Clark, B.~R. Miller, B.V. Saunders,
                H.S. Cohl, and M.A. McClain, eds."}

@article{Leaver:1985ax,
    author = "Leaver, E. W.",
    title = "{An Analytic representation for the quasi normal modes of Kerr black holes}",
    doi = "10.1098/rspa.1985.0119",
    journal = "Proc. Roy. Soc. Lond. A",
    volume = "402",
    pages = "285--298",
    year = "1985"
}

@article{Leaver:1986gd,
    author = "Leaver, Edward W.",
    title = "{Spectral decomposition of the perturbation response of the Schwarzschild geometry}",
    doi = "10.1103/PhysRevD.34.384",
    journal = "Phys. Rev. D",
    volume = "34",
    pages = "384--408",
    year = "1986"
}

@article{Andersson:1996cm,
    author = "Andersson, Nils",
    title = "{Evolving test fields in a black hole geometry}",
    eprint = "gr-qc/9607064",
    archivePrefix = "arXiv",
    reportNumber = "WUGRAV-96-6, PREPRINT-NO-WUGRAV-96-6",
    doi = "10.1103/PhysRevD.55.468",
    journal = "Phys. Rev. D",
    volume = "55",
    pages = "468--479",
    year = "1997"
}

@article{DeAmicis:2024not,
    author = "De Amicis, Marina and Albanesi, Simone and Carullo, Gregorio",
    title = "{Inspiral-inherited ringdown tails}",
    eprint = "2406.17018",
    archivePrefix = "arXiv",
    primaryClass = "gr-qc",
    doi = "10.1103/PhysRevD.110.104005",
    journal = "Phys. Rev. D",
    volume = "110",
    number = "10",
    pages = "104005",
    year = "2024"
}

@article{DeAmicis:2025xuh,
    author = "De Amicis, Marina and Cannizzaro, Enrico and Carullo, Gregorio and Sberna, Laura",
    title = "{Dynamical quasinormal mode excitation}",
    eprint = "2506.21668",
    archivePrefix = "arXiv",
    primaryClass = "gr-qc",
    month = "6",
    year = "2025",
    journal=""
}

@book{abramowitz1968handbook,
  title={Handbook of mathematical functions},
  author={Abramowitz, Milton and Stegun, Irene A},
  volume={55},
  year={1968},
  publisher={US Government printing office}
}

@article{Casals:2015nja,
    author = "Casals, Marc and Ottewill, Adrian C.",
    title = "{High-order tail in Schwarzschild spacetime}",
    eprint = "1509.04702",
    archivePrefix = "arXiv",
    primaryClass = "gr-qc",
    doi = "10.1103/PhysRevD.92.124055",
    journal = "Phys. Rev. D",
    volume = "92",
    number = "12",
    pages = "124055",
    year = "2015"
}

@article{DeAmicis:2024eoy,
    author = "De Amicis, Marina and others",
    title = "{Late-Time Tails in Nonlinear Evolutions of Merging Black Holes}",
    eprint = "2412.06887",
    archivePrefix = "arXiv",
    primaryClass = "gr-qc",
    doi = "10.1103/2brx-xnyr",
    journal = "Phys. Rev. Lett.",
    volume = "135",
    number = "17",
    pages = "171401",
    year = "2025"
}

@article{Nagar:2005ea,
    author = "Nagar, Alessandro and Rezzolla, Luciano",
    title = "{Gauge-invariant non-spherical metric perturbations of Schwarzschild black-hole spacetimes}",
    eprint = "gr-qc/0502064",
    archivePrefix = "arXiv",
    doi = "10.1088/0264-9381/22/16/R01",
    journal = "Class. Quant. Grav.",
    volume = "22",
    pages = "R167",
    year = "2005",
    note = "[Erratum: Class.Quant.Grav. 23, 4297 (2006)]"
}

@article{Pound:2021qin,
    author = "Pound, Adam and Wardell, Barry",
    title = "{Black hole perturbation theory and gravitational self-force}",
    eprint = "2101.04592",
    archivePrefix = "arXiv",
    primaryClass = "gr-qc",
    month = "1",
    year = "2021",
    journal=""
}

@article{Buonanno:1998gg,
    author = "Buonanno, A. and Damour, T.",
    title = "{Effective one-body approach to general relativistic two-body dynamics}",
    eprint = "gr-qc/9811091",
    archivePrefix = "arXiv",
    reportNumber = "IHES-P-98-74",
    doi = "10.1103/PhysRevD.59.084006",
    journal = "Phys. Rev. D",
    volume = "59",
    pages = "084006",
    year = "1999"
}

@article{Damour:2009zoi,
    author = "Damour, Thibault and Nagar, Alessandro",
    editor = "Blanchet, Luc and Spallicci, Alessandro and Whiting, Bernard",
    title = "{The Effective One Body description of the Two-Body problem}",
    eprint = "0906.1769",
    archivePrefix = "arXiv",
    primaryClass = "gr-qc",
    doi = "10.1007/978-90-481-3015-3_7",
    journal = "Fundam. Theor. Phys.",
    volume = "162",
    pages = "211--252",
    year = "2011"
}

@article{London:2014cma,
    author = "London, Lionel and Shoemaker, Deirdre and Healy, James",
    title = "{Modeling ringdown: Beyond the fundamental quasinormal modes}",
    eprint = "1404.3197",
    archivePrefix = "arXiv",
    primaryClass = "gr-qc",
    doi = "10.1103/PhysRevD.90.124032",
    journal = "Phys. Rev. D",
    volume = "90",
    number = "12",
    pages = "124032",
    year = "2014",
    note = "[Erratum: Phys.Rev.D 94, 069902 (2016)]"
}

@article{London:2018gaq,
    author = "London, L. T.",
    title = "{Modeling ringdown. II. Aligned-spin binary black holes, implications for data analysis and fundamental theory}",
    eprint = "1801.08208",
    archivePrefix = "arXiv",
    primaryClass = "gr-qc",
    doi = "10.1103/PhysRevD.102.084052",
    journal = "Phys. Rev. D",
    volume = "102",
    number = "8",
    pages = "084052",
    year = "2020"
}

@article{Baker:2008mj,
    author = "Baker, John G. and Boggs, William D. and Centrella, Joan and Kelly, Bernard J. and McWilliams, Sean T. and van Meter, James R.",
    title = "{Mergers of non-spinning black-hole binaries: Gravitational radiation characteristics}",
    eprint = "0805.1428",
    archivePrefix = "arXiv",
    primaryClass = "gr-qc",
    doi = "10.1103/PhysRevD.78.044046",
    journal = "Phys. Rev. D",
    volume = "78",
    pages = "044046",
    year = "2008"
}

@article{Damour:2014yha,
    author = "Damour, Thibault and Nagar, Alessandro",
    title = "{A new analytic representation of the ringdown waveform of coalescing spinning black hole binaries}",
    eprint = "1406.0401",
    archivePrefix = "arXiv",
    primaryClass = "gr-qc",
    doi = "10.1103/PhysRevD.90.024054",
    journal = "Phys. Rev. D",
    volume = "90",
    number = "2",
    pages = "024054",
    year = "2014"
}

@article{Bonga:2018zlx,
    author = "Bonga, Beatrice and Poisson, Eric and Yang, Huan",
    title = "{Self-torque and angular momentum balance for a spinning charged sphere}",
    eprint = "1805.01372",
    archivePrefix = "arXiv",
    primaryClass = "physics.class-ph",
    doi = "10.1119/1.5054590",
    journal = "Am. J. Phys.",
    volume = "86",
    number = "11",
    pages = "839",
    year = "2018"
}

@article{Berti:2015itd,
    author = "Berti, Emanuele and others",
    title = "{Testing General Relativity with Present and Future Astrophysical Observations}",
    eprint = "1501.07274",
    archivePrefix = "arXiv",
    primaryClass = "gr-qc",
    doi = "10.1088/0264-9381/32/24/243001",
    journal = "Class. Quant. Grav.",
    volume = "32",
    pages = "243001",
    year = "2015"
}

@book{10.1093/acprof:oso/9780199205677.001.0001,
    author = {Alcubierre, Miguel},
    title = {Introduction to 3+1 Numerical Relativity},
    publisher = {Oxford University Press},
    year = {2008},
    month = {04},
    isbn = {9780199205677},
    doi = "{10.1093/acprof:oso/9780199205677.001.0001}",
    url = {https://doi.org/10.1093/acprof:oso/9780199205677.001.0001},
}

@article{Price:1971fb,
    author = "Price, Richard H.",
    title = "{Nonspherical perturbations of relativistic gravitational collapse. 1. Scalar and gravitational perturbations}",
    doi = "10.1103/PhysRevD.5.2419",
    journal = "Phys. Rev. D",
    volume = "5",
    pages = "2419--2438",
    year = "1972"
}

@article{Oshita:2025qmn,
    author = "Oshita, Naritaka and Ma, Sizheng and Chen, Yanbei and Yang, Huan",
    title = "{Probing Direct Waves in Black Hole Ringdowns}",
    journal = "arXiv",
    eprint = "2509.09165",
    archivePrefix = "arXiv",
    primaryClass = "gr-qc",
    year = "2025"
}

@article{Bernuzzi:2010ty,
    author = "Bernuzzi, Sebastiano and Nagar, Alessandro",
    title = "{Binary black hole merger in the extreme-mass-ratio limit: a multipolar analysis}",
    eprint = "1003.0597",
    archivePrefix = "arXiv",
    primaryClass = "gr-qc",
    doi = "10.1103/PhysRevD.81.084056",
    journal = "Phys. Rev. D",
    volume = "81",
    pages = "084056",
    year = "2010"
}

@article{Bernuzzi:2011aj,
    author = "Bernuzzi, Sebastiano and Nagar, Alessandro and Zenginoglu, Anil",
    title = "{Binary black hole coalescence in the large-mass-ratio limit: the hyperboloidal layer method and waveforms at null infinity}",
    eprint = "1107.5402",
    archivePrefix = "arXiv",
    primaryClass = "gr-qc",
    doi = "10.1103/PhysRevD.84.084026",
    journal = "Phys. Rev. D",
    volume = "84",
    pages = "084026",
    year = "2011"
}

@article{LIGOScientific:2025csr,
    author = "Abac, A. G. and others",
    collaboration = "LIGO Scientific, VIRGO, KAGRA",
    title = "{Directed searches for gravitational waves from ultralight vector boson clouds around merger remnant and galactic black holes during the first part of the fourth LIGO-Virgo-KAGRA observing run}",
    eprint = "2509.07352",
    archivePrefix = "arXiv",
    primaryClass = "gr-qc",
    reportNumber = "LIGO-P2500256",
    month = "9",
    year = "2025",
    journal=""
}

@article{Mano:1996vt,
    author = "Mano, Shuhei and Suzuki, Hisao and Takasugi, Eiichi",
    title = "{Analytic solutions of the Teukolsky equation and their low frequency expansions}",
    eprint = "gr-qc/9603020",
    archivePrefix = "arXiv",
    reportNumber = "OU-HET-238",
    doi = "10.1143/PTP.95.1079",
    journal = "Prog. Theor. Phys.",
    volume = "95",
    pages = "1079--1096",
    year = "1996"
}

@article{Mano:1996mf,
    author = "Mano, Shuhei and Suzuki, Hisao and Takasugi, Eiichi",
    title = "{Analytic solutions of the Regge-Wheeler equation and the postMinkowskian expansion}",
    eprint = "gr-qc/9605057",
    archivePrefix = "arXiv",
    reportNumber = "OU-HET-246",
    doi = "10.1143/PTP.96.549",
    journal = "Prog. Theor. Phys.",
    volume = "96",
    pages = "549--566",
    year = "1996"
}

@article{Sasaki:2003xr,
    author = "Sasaki, Misao and Tagoshi, Hideyuki",
    title = "{Analytic black hole perturbation approach to gravitational radiation}",
    eprint = "gr-qc/0306120",
    archivePrefix = "arXiv",
    doi = "10.12942/lrr-2003-6",
    journal = "Living Rev. Rel.",
    volume = "6",
    pages = "6",
    year = "2003"
}

@article{LIGOScientific:2025hdt,
    author = "Abac, A. G. and others",
    collaboration = "LIGO Scientific, KAGRA, VIRGO",
    title = "{GWTC-4.0: An Introduction to Version 4.0 of the Gravitational-Wave Transient Catalog}",
    eprint = "2508.18080",
    archivePrefix = "arXiv",
    primaryClass = "gr-qc",
    reportNumber = "LIGO-P2400293",
    doi = "10.3847/2041-8213/ae0c06",
    journal = "Astrophys. J. Lett.",
    volume = "995",
    number = "1",
    pages = "L18",
    year = "2025"
}

@article{Saleem:2021iwi,
    author = "Saleem, M. and others",
    title = "{The science case for LIGO-India}",
    eprint = "2105.01716",
    archivePrefix = "arXiv",
    primaryClass = "gr-qc",
    doi = "10.1088/1361-6382/ac3b99",
    journal = "Class. Quant. Grav.",
    volume = "39",
    number = "2",
    pages = "025004",
    year = "2022"
}

@article{Pandey:2024mlo,
    author = "Pandey, Shiksha and Gupta, Ish and Chandra, Koustav and Sathyaprakash, Bangalore S.",
    title = "{The Critical Role of LIGO-India in the Era of Next-generation Observatories}",
    eprint = "2411.10349",
    archivePrefix = "arXiv",
    primaryClass = "gr-qc",
    doi = "10.3847/2041-8213/add15f",
    journal = "Astrophys. J. Lett.",
    volume = "985",
    number = "1",
    pages = "L17",
    year = "2025"
}

@article{Colpi:2024xhw,
    author = "Colpi, Monica and others",
    title = "{LISA Definition Study Report}",
    eprint = "2402.07571",
    archivePrefix = "arXiv",
    primaryClass = "astro-ph.CO",
    month = "2",
    year = "2024",
    journal="",
}

@article{Abac:2025saz,
    author = "Abac, Adrian and others",
    title = "{The Science of the Einstein Telescope}",
    eprint = "2503.12263",
    archivePrefix = "arXiv",
    primaryClass = "gr-qc",
    reportNumber = "ET-0036C-25",
    month = "3",
    year = "2025",
    journal="",
}

@article{Arnaudo:2025uos,
    author = "Arnaudo, Paolo and Carballo, Javier and Withers, Benjamin",
    title = "{Beyond quasinormal modes: a complete mode decomposition of black hole perturbations}",
    eprint = "2510.18956",
    archivePrefix = "arXiv",
    primaryClass = "gr-qc",
    month = "10",
    year = "2025",
    journal=""
}

@article{Kuntz:2025gdq,
    author = "Kuntz, Adrien",
    title = "{Green function of the Pöschl-Teller potential}",
    eprint = "2510.17954",
    archivePrefix = "arXiv",
    primaryClass = "gr-qc",
    month = "10",
    year = "2025",
    journal=""
}

@article{LIGOScientific:2021sio,
    author = "Abbott, R. and others",
    collaboration = "LIGO Scientific, VIRGO, KAGRA",
    title = "{Tests of General Relativity with GWTC-3}",
    eprint = "2112.06861",
    archivePrefix = "arXiv",
    primaryClass = "gr-qc",
    reportNumber = "LIGO-P2100275",
    doi = "10.1103/PhysRevD.112.084080",
    journal = "Phys. Rev. D",
    volume = "112",
    number = "8",
    pages = "084080",
    year = "2025"
}

@article{Leaver:1986vnb,
    author = "Leaver, E. W.",
    title = "{Solutions to a generalized spheroidal wave equation: Teukolsky's equations in general relativity, and the two-center problem in molecular quantum mechanics}",
    doi = "10.1063/1.527130",
    journal = "J. Math. Phys.",
    volume = "27",
    number = "5",
    pages = "1238",
    year = "1986"
}

@article{LIGOScientific:2025rsn,
    author = "Abac, A. G. and others",
    collaboration = "LIGO Scientific, VIRGO, KAGRA",
    title = "{GW231123: A Binary Black Hole Merger with Total Mass 190{$\textendash$}265 $M_{\odot}$}",
    eprint = "2507.08219",
    archivePrefix = "arXiv",
    primaryClass = "astro-ph.HE",
    reportNumber = "DCC: P2500026-v6, DCC: P2500026-v8",
    doi = "10.3847/2041-8213/ae0c9c",
    journal = "Astrophys. J. Lett.",
    volume = "993",
    number = "1",
    pages = "L25",
    year = "2025"
}

@article{LIGOScientific:2025rid,
    author = "Abac, A. G. and others",
    collaboration = "LIGO Scientific, Virgo, KAGRA",
    title = "{GW250114: Testing Hawking's Area Law and the Kerr Nature of Black Holes}",
    eprint = "2509.08054",
    archivePrefix = "arXiv",
    primaryClass = "gr-qc",
    reportNumber = "LIGO-P2500421",
    doi = "10.1103/kw5g-d732",
    journal = "Phys. Rev. Lett.",
    volume = "135",
    number = "11",
    pages = "111403",
    year = "2025"
}

@article{LIGOScientific:2025brd,
    author = "Abac, A. G. and others",
    collaboration = "LIGO Scientific, Virgo, KAGRA",
    title = "{GW241011 and GW241110: Exploring Binary Formation and Fundamental Physics with Asymmetric, High-spin Black Hole Coalescences}",
    eprint = "2510.26931",
    archivePrefix = "arXiv",
    primaryClass = "astro-ph.HE",
    reportNumber = "LIGO-P2500402",
    doi = "10.3847/2041-8213/ae0d54",
    journal = "Astrophys. J. Lett.",
    volume = "993",
    number = "1",
    pages = "L21",
    year = "2025"
}

@article{Arnaudo:2025kit,
    author = "Arnaudo, Paolo and Withers, Benjamin",
    title = "{Price's law from quasinormal modes}",
    eprint = "2511.17703",
    archivePrefix = "arXiv",
    primaryClass = "gr-qc",
    month = "11",
    year = "2025",
    journal=""
}

@article{LIGOScientific:2014pky,
    author = "Aasi, J. and others",
    collaboration = "LIGO Scientific",
    title = "{Advanced LIGO}",
    eprint = "1411.4547",
    archivePrefix = "arXiv",
    primaryClass = "gr-qc",
    doi = "10.1088/0264-9381/32/7/074001",
    journal = "Class. Quant. Grav.",
    volume = "32",
    pages = "074001",
    year = "2015"
}

@article{VIRGO:2014yos,
    author = "Acernese, F. and others",
    collaboration = "VIRGO",
    title = "{Advanced Virgo: a second-generation interferometric gravitational wave detector}",
    eprint = "1408.3978",
    archivePrefix = "arXiv",
    primaryClass = "gr-qc",
    doi = "10.1088/0264-9381/32/2/024001",
    journal = "Class. Quant. Grav.",
    volume = "32",
    number = "2",
    pages = "024001",
    year = "2015"
}

@article{KAGRA:2020tym,
    author = "Akutsu, T. and others",
    collaboration = "KAGRA",
    title = "{Overview of KAGRA: Detector design and construction history}",
    eprint = "2005.05574",
    archivePrefix = "arXiv",
    primaryClass = "physics.ins-det",
    doi = "10.1093/ptep/ptaa125",
    journal = "PTEP",
    volume = "2021",
    number = "5",
    pages = "05A101",
    year = "2021"
}

\end{document}